\documentclass[twocolumn]{aastex631}

\usepackage{enumitem}
\usepackage{hyperref}
\usepackage{booktabs}
\usepackage{tabularx}
\usepackage[table]{xcolor}
\usepackage{amsmath}
\usepackage{mathdots}
\usepackage{multirow}
\usepackage{mathrsfs}
\usepackage{float}

\usepackage{bm}
\renewcommand{\vec}[1]{\bm{#1}}

\begin{document}

% \title{Efficient parameterization of ion velocity distribution functions using Slepian functions}
\title{Recovering Ion Distribution Functions: \\
II. Gyrotropic Slepian Reconstruction of Solar Wind Electrostatic Analyzer Measurements}

\author[0000-0003-0896-7972]{Srijan Bharati Das}
\affiliation{Center for Astrophysics | Harvard \& Smithsonian, 
60 Garden Street, Cambridge, MA 02138, USA.}
\email{srijanbdas@alumni.princeton.edu}

\author[0000-0003-4747-6252]{Michael Terres}
\affiliation{Center for Astrophysics | Harvard \& Smithsonian, 
60 Garden Street, Cambridge, MA 02138, USA.}
\email{michael.terres.sci@gmail.com}

\shorttitle{Slepian reconstruction of gyrotropic ion VDFs}
\shortauthors{Bharati Das \& Terres}



\begin{abstract}

Velocity distribution functions (VDF) are an essential observable for studying kinetic and wave-particle processes in solar wind plasmas. To experimentally distinguish modes of heating, acceleration, and turbulence in the solar wind, precise representations of particle phase space VDFs are needed. In the first paper of this series,  we developed the Slepian Basis Reconstruction (SBR) method to approximate fully agyrotropic continuous distributions from discrete measurements of electrostatic analyzers (ESAs). The method enables accurate determination of plasma moments, preserves kinetic features, and prescribes smooth gradients in phase space. In this paper, we extend the SBR method by imposing gyrotropic symmetry (g-SBR). Incorporating this symmetry enables high-fidelity reconstruction of VDFs that are partially measured, as from an ESA with a limited field-of-view (FOV). We introduce three frameworks for g-SBR, the gyrotropic Slepian Basis Reconstruction: (A) 1D angular Slepian functions on a polar-cap, (B) 2D Slepian functions in a Cartesian plane, and (C) a hybrid method. We employ model distributions representing multiple anisotropic ion populations in the solar wind to benchmark these methods, and we show that the g-SBR method produces a reconstruction that preserves kinetic structures and plasma moments, even with a strongly limited FOV. For our choice of model distribution, g-SBR can recover $\geq90\%$ of the density when only $20\%$ is measured. We provide the package \texttt{gdf} for open-source use and contribution by the heliophysics community. This work establishes direct pathways to bridge particle observations with kinetic theory and simulations, facilitating the investigation of gyrotropic plasma heating phenomena across the heliosphere.

\end{abstract}



% \begin{itemize}
%     \item Introduction
%     \item Methods
%     \begin{itemize}
%         \item Finding Axis of Gyrotopy [Fig 1]
%         \item Slepian Reconstruction [Fig 2, Fig 3]
%         \item In Preparation for FOV Restricted ESAs [Fig 4]
%     \end{itemize}
%     \item Final Remarks [Fig 5]
%     \item Appendix
%     \begin{itemize}
%         \item Math
%         \item Slepian on Polar Cap
%         \item Electron VDF with Spherical Harmonics: Comparison to Vinas et al. 2009.
%         \itme Ion VDF with Spherical Harmonics: Motivation for Slepian Functions
%     \end{itemize}
% \end{itemize}

\section{Introduction} \label{sec:intro}

Transport phenomena in the Sun's corona and interplanetary space leave distinct and often distinguishable signatures in the velocity distribution functions of solar wind ions (VDFs). In the turbulent, weakly collisional solar wind, Coulomb collisions are insufficient to drive VDFs to local thermodynamic equilibrium (LTE). Wave-particle interactions and plasma instabilities (magnetohydrodynamic and kinetic) serve as sources of free energy, producing often complex non-LTE distributions \citep{verscharen2019multi}. These non-Maxwellian features, such as temperature anisotropies, interpenetrating beams, and non-thermal tails, signify different modes of prior interaction and different susceptibilities to ongoing interaction in the plasma. 

Proton distribution functions, which are considerably distorted by underlying wave-particle interactions, serve as a valuable tool for comprehending the processes through which turbulence dissipates in the solar wind. Landau damping \citep{Shankarappa_2023}, ion cyclotron resonance \citep{Bowen_2024, Terres_2022}, stochastic heating \citep{Martinovic_2020, Chandran_2013}, and magnetic reconnection \citep{Vech_2018,Terres_2022}, for example, are all processes by which turbulent fluctuations are dissipated into the particle populations. Each of these processes embeds characteristic signatures in the VDF, and the relative importance of these often inter-operating mechanisms is an area of active research \citep[for a recent review of plasma turbulence and dissipation see][]{Howes_2024}. To address the key open questions, complete and well-resolved measurements of phase space proton VDFs are essential. 

Electrostatic analyzers (ESAs) sample ion VDFs across discrete energy and angular bins \citep{Fazakerley_Measurements}. Gyrotropy is often a reasonable approximation when the accumulation time of the ESA is much larger than the ion gyro-period (inversely proportional to the magnetic field strength). For such measurements, gyrotropic binning or regression to a gyrotropic model may reveal coherent kinetic features and reduce noise. When the measurement geometry is partially obscured by an obstacle (such as in SPAN-Ai onboard Parker Solar Probe \citep{Livi_etal_2022}) or limited in angular extent, portions of the distribution are missing. Under these circumstances, gyrotropy can provide the required symmetry to infer the unmeasured portion of phase space.
Common parametric fitting approaches, such as regression to bi-Maxwellians, are inadequate to capture fine-scale kinetic structures, non-Gaussian peak shapes, and other signatures. Furthermore, the iterative methods typically employed for fitting such models exhibit sensitivity to initial estimates and may fail to converge within under-determined or masked measurement regimes \citep{Verniero_2020, Klein_2021}.

To address these challenges, we extend the fully agyrotropic Slepian Basis Reconstruction (SBR) methodology developed by \citet{Das_Terres_2025} to gyrotropic ion VDFs, inspired by SPAN-Ai measurements. Slepian functions form a family of orthogonal basis functions that are simultaneously localized in spatial and spectral domains, making them uniquely suited for representing compact structures in velocity space. When adapted to exploit the gyrotropic symmetry of the solar wind, anchored around the magnetic field direction, Slepian functions can form a reduced representation of the VDF with minimal loss of information. This makes the SBR method especially powerful for reconstructing distributions from partial observations, as is the case with SPAN-Ai measurements when outside its primary mission objective (i.e., Parker Solar Probes tangential speed is low).

\section{Methodology} \label{sec:methods}

In this section, we introduce the method for reconstructing gyrotropic distribution functions (GDF) and the accompanying Python package, \texttt{gdf}\footnote{\href{https://srijaniiserprinceton.github.io/GDF/index.html}{https://srijaniiserprinceton.github.io/GDF}}. Throughout the paper, GDF refers to the reconstruction method in general, while \texttt{gdf} (in stylized text) refers specifically to the software implementation.

\begin{figure}
    \centering
    \includegraphics[width=\linewidth]{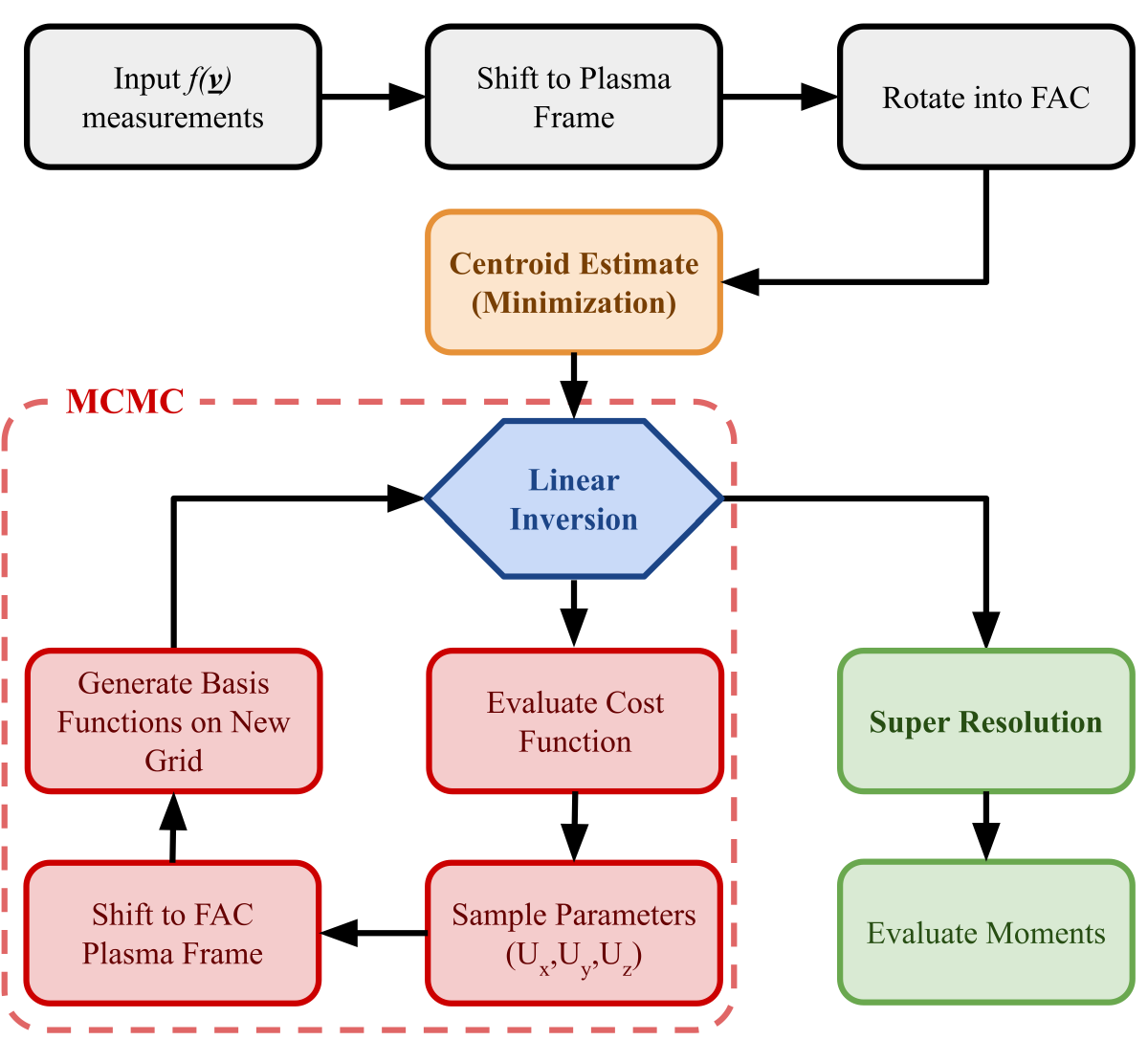}
    \caption{Flow diagram illustrating the steps starting from the ESA measured 3D VDF leading to our GDF reconstruction and final moment calculation.}
    \label{fig:Flowdiagram}
\end{figure}

\begin{figure*}
    \centering
    \includegraphics[width=\linewidth]{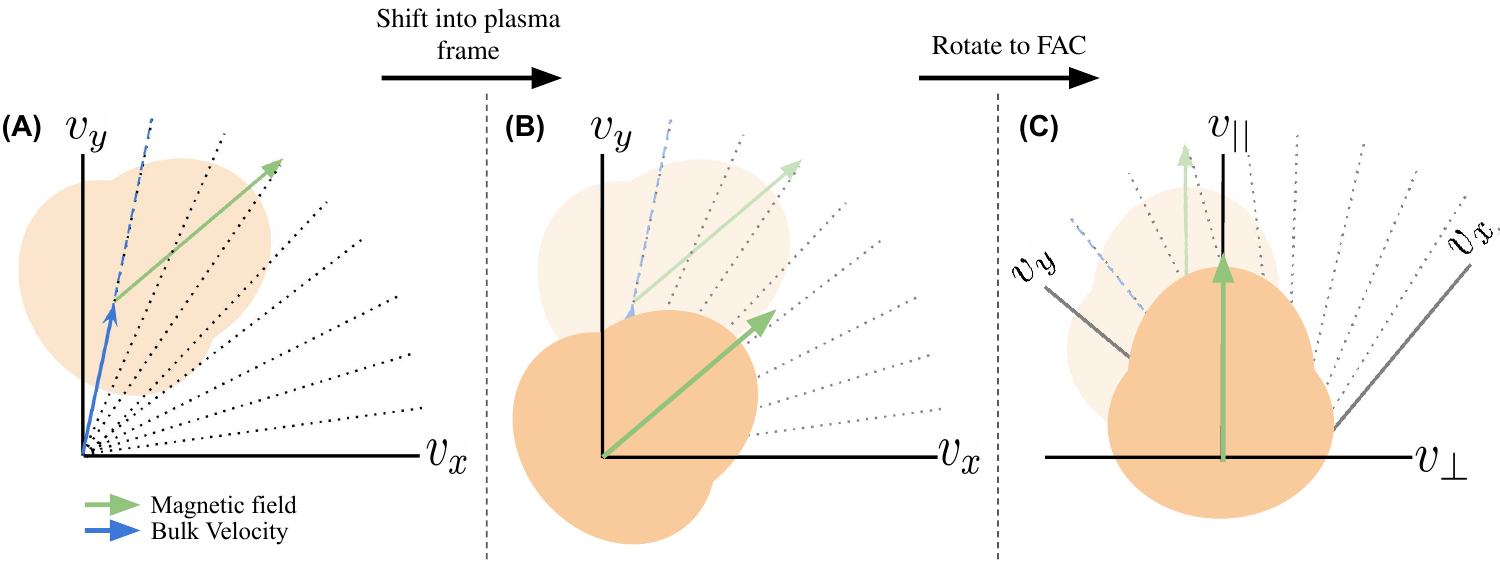}
    \caption{Schematic figure showing the pre-processing steps before fitting the VDF using the gyrotropic model in Eqn.~(\ref{eqn:f_gyro_1DSlep}). Panel (A) represents the VDF in the instrument frame before any pre-processing, the example magnetic field vector is entirely contained in the $v_x-v_y$ plane. For simplicity, we only show the FOV restricted ESA grids (using dots) in a single plane of elevation. Panel (B) demonstrates the shift of the VDF to the plasma frame for a certain bulk velocity $\boldsymbol{U}$. Panel (C) shows the final magnetic field aligned VDF, resulting in the $(v_{\parallel}, v_{\perp})$ axes. The bulk velocity vector in the instrument frame is shown with the blue arrow and the magnetic field direction is shown using the green arrow.}
    \label{fig:FAC}
\end{figure*}

\begin{figure*}
    \centering
    \includegraphics[width=\linewidth]{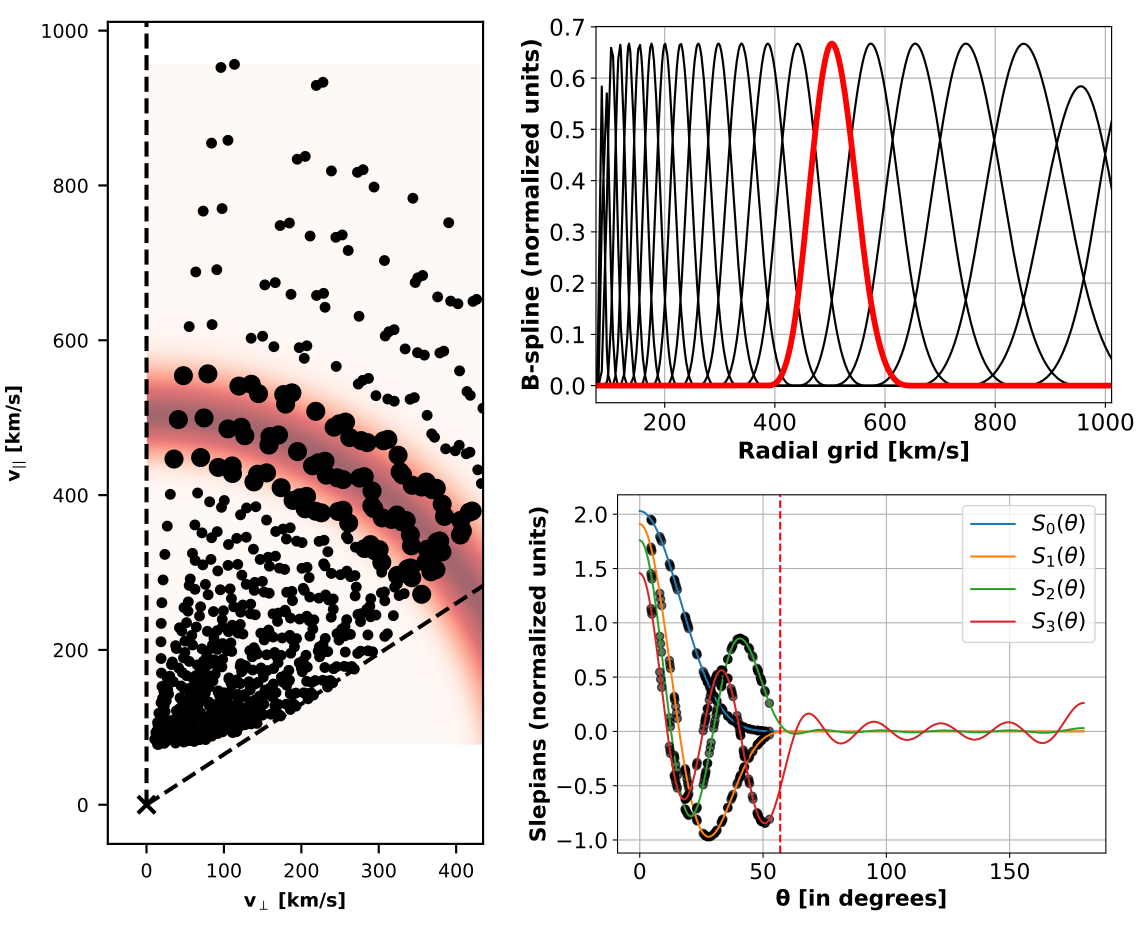}
    \caption{Fitting architecture when using 1D Slepians on a polar cap. Panel (A) shows the distribution of grid points in black after rotating to FAC followed by boosting the frame to induce a maximum angle of $\Theta$ to the grid points about the origin. The series of operations of obtain the FAC is demonstrated in Fig.~\ref{fig:FAC} and the frame boosting is performed according to Eqn.~(\ref{eqn:frame_boost}). The set of B-splines corresponding to the synthetic ESA grids projected into FAC is shown in panel (B). The local support in radius for the \textit{red} B-spline highlighted is shown as the shaded red arc in panel (A). All grid points encompassed by this B-spline are marked with a larger size. Panel (C) shows the Slepian functions optimally concentrated within angle $\Theta$ about the gyroaxis. The distribution of the grid points within the red arc of panel (A) is overplotted on each of the Slepian functions $S_{\alpha}(\theta)$. The vertical \textit{red dashed} line marks $\Theta$ used for the polar cap shown in panel (A).}
    \label{fig:Fitting_Setup}
\end{figure*}

The method for reconstructing gyrotropic distribution functions (GDF) is motivated by (I) a desire for robust continuum gyrotropic reconstructions from discrete ESA measurements, and (II) the challenges presented by field-of-view–obstructed ESA measurements, such as those encountered by the SPAN-Ai instrument \citep{Livi_etal_2022} onboard PSP. While not utilizing SPAN-Ai data or techniques directly, our approach addresses similar measurement constraints by employing a coarsely gridded, partial velocity space domain to generate Slepian basis coefficients. In this paper, all of the example cases are based on \href{http://sweap.cfa.harvard.edu}{SWEAP} science scenarios. The resulting reconstruction serves as a Slepian-based generative model, producing continuous distributions in the gyrotropic, field-aligned velocity space.

The primary assumption that goes into GDF is that solar wind particle distribution functions are gyrotropic. For typical solar-wind conditions, the ESA accumulation time is much longer than the proton and electron gyro-periods. As a result, the measured velocity distribution functions are averaged over many cyclotron orbits, thus removing any gyro-phase dependence. This assumption holds regardless of the plasma beta, $\beta = n k_b T/(B^2/2\mu_0)$, because even a high-$\beta$ plasma can satisfy the gyrotropic condition as long as the ion Larmor radius is small compared to the macroscopic gradients of the equilibrium plasma \citep{Howes_2006}.

The algorithm for reconstructing smooth GDFs is summarized in Fig.~\ref{fig:Flowdiagram}. It begins with loading the measured VDF on instrument-defined grids. Using grid-integrated moments, the distribution is transformed into the plasma frame, then rotated into the field-aligned coordinate system.  The algorithm then proceeds in two stages: (A) gyro-centroid finding, and (B) super-resolution. The gyrocenter is determined via a Markov Chain Monte Carlo (MCMC) inversion (Section~\ref{sec:gyro-centroid-finder}). The super-resolution process is performed by one of the three GDF generation methods discussed later in this section. From the super-resolved VDF, moments can then be evaluated directly.

\subsection{Gyro-centroid finder} \label{sec:gyro-centroid-finder}
%The primary assumption that goes into reconstructing our GDFs is that solar wind particle distribution functions are gyrotropic. ESA measurements have a finite time cadence. For example, the SPAN-Ai measurement cadence varies through an encounter, with the prime mission phase collection time ranging between 3.5 seconds to 1.7 seconds (\textbf{put exact numbers here}). 
An ESA measures particle flux by sequentially sampling different energies and angles over a fixed accumulation time. During the accumulation period, it counts particles that pass through its energy–angle selection, producing a time-averaged estimate of the velocity distribution function for each measurement bin \citep{verscharen2019multi}. Since a gyrotropic distribution is symmetric about the magnetic field direction, one requires an estimate for the magnetic unit vector $\hat{\boldsymbol{b}}$ during this accumulation period. Most space-based ESAs are paired with magnetometers that sample at high frequency relative to the accumulation time, so for this algorithm, we assume that $\hat{\boldsymbol{b}}$ is known. This provides the orientation of the gyrosymmetry axis (gyroaxis), however the gyroaxis must also be chosen so as to cross the centroid of the VDF. Locating the gyroaxis may be particularly challenging if the instrument grid does not thoroughly sample a velocity space volume that includes the centroid. 

To determine the location of the gyroaxis for a given set of measurements, we take a joint inversion approach that identifies both the gyroaxis position and the gyrotropic distribution that best fits the data. The model becomes simpler when the instrument grids are re-expressed in the magnetic ield-aligned coordinates, (FAC), reducing it to a ``2.5D" mathematical framework where the 3D distribution is symmetrical about the magnetic axis. Thus, we transition from the $(v_x, v_y, v_z)$ instrument grids to the $(v_{\parallel}, v_{\perp 1}, v_{\perp 2})$ FAC convention. We use 1D cubic B-splines in velocity and 1D Slepian functions in polar angle to compose our 2.5D model as below.
\begin{equation} \label{eqn:f_gyro_1DSlep}
    \log_{10}{f_{\rm{gyro}}(v_{\parallel}, v_{\perp})} = \sum_{\alpha=0}^{N^{\mathrm{2D}}_{\mathrm{polcap}}} \sum_{i=0}^{N} c^{i,\alpha} \, \beta_i(v) \, g_{\alpha}^{\mathrm{polcap}}(\theta) \\ .
\end{equation}
Here, $v = \sqrt{v_{\parallel}^2 + v_{\perp}^2}$, $v_{\perp} = \sqrt{v_{\perp 1}^2 + v_{\perp 2}^2}$ and $\theta = \tan^{-1}(v_{\perp} / v_{\parallel})$ is the elevation angle ranging between $[-90^{\circ},90^{\circ}]$ where $\theta = 0^{\circ}$ is aligned along the gyroaxis. We refer the reader to Sections~3 through 5 of \cite{Slepian_polar_caps} for the complete theoretical description of generating Slepians on a polar cap $g^{\mathrm{polcap}}_{\alpha}(\theta)$. The polar cap Slepians are a sub-class of Slepian functions generated on the surface of a sphere. In our case, we only use the polar cap Slepian functions which are symmetric about the polar axis (in this case $\hat{\boldsymbol{b}}$). As shown in \cite{Slepian_polar_caps}, the polar cap Slepian basis is calculated numerically and are the eigenfunctions satisfying a Fredholm integral eigenvalue equation. The polar cap Slepian basis depends on two input parameters --- the angular extent $\Theta$ of the polar cap about $\hat{\boldsymbol{b}}$ and the maximum angular degree of the basis functions $L_{\mathrm{max}}$.

The \textsc{Matlab} software used in this paper to generate the polar-cap Slepian basis can be found in Zenodo references \cite{slepian_alpha} and \cite{slepian_foxtrot}. In particular, we used the \textsc{Matlab} function \texttt{sdwcap.m} under the \textsc{slepian\_alpha} repository. It is important to note that the field-aligned coordinates \((v_{\parallel}, v_{\perp 1}, v_{\perp 2})\) depend on the choice of bulk velocity \(\boldsymbol{U} = (U_x, U_y, U_z)\), which is used to shift to the plasma frame before rotating the coordinates to align with the magnetic field direction \(\hat{\boldsymbol{b}}\). While the final bulk velocity moments are computed from the reconstructed VDF, it is evident that the selected bulk velocity influences the recovered gyrotropic VDF. Therefore, we perform a joint optimization that involves (a) inferring the most likely gyrotropic axis and (b) reconstructing a gyrotropic VDF that minimizes the misfit compared to ESA measurements. Inferring the gyroaxis requires a prescription of how we choose to model the gyrotropic distribution. Therefore, although the first step in our method is to determine the gyroaxis, we first describe our model parameterization before detailing the gyroaxis estimation (see Sec.~\ref{sec:gyroaxis_estimation}). 

After shifting to the plasma frame using $\boldsymbol{U}$ and rotating to the field aligned coordinates using $\hat{\boldsymbol{b}}$, we obtain the re-mapped instrument grids in the $(v_{\parallel}, v_{\perp 1}, v_{\perp 2})$ coordinate frame. This series of steps is demonstrated schematically in Fig.~\ref{fig:FAC}. Naturally, this coordinate frame is anchored with the bulk velocity $\boldsymbol{U}$ being the origin. To perform VDF gyrotropization using polar Slepian functions, we need the VDF to exist on localized polar caps around the origin of the coordinate frame. However, in the plasma frame, the origin is at the bulk velocity, and the VDF would span the full $4\pi$ angular extent around the origin. Therefore, we boost all the grids along the parallel direction such that $v_{\parallel} \rightarrow v_{\parallel} - \Delta v_{\parallel}$. This boost magnitude $\Delta v_{\parallel}$ is chosen such that
\begin{eqnarray} \label{eqn:frame_boost}
    \Delta v_{\parallel} &=& v_{\mathrm{inst, max-val}} - v_{\parallel, \mathrm{max-val}}, \, \\
    v_{\parallel} &\rightarrow& v_{\parallel} - \Delta v_{\parallel} \, ,
\end{eqnarray}
where $v_{\mathrm{inst, max-val}}$ represents the velocity magnitude in the instrument frame for the peak VDF value and $v_{\parallel, \mathrm{max-val}}$ is the parallel velocity for the grid corresponding to the peak VDF value in the FAC frame. The above boosting of the parallel velocity coordinate ensures that the distance of the peak VDF value from the origin of this boosted FAC is the same as its distance from the origin in the instrument frame. 

The polar cap of angular extent $\Theta$ on which we generate our Slepian functions is chosen to be the maximum elevation $\theta$ that a grid point subtends about the new origin. This polar cap angular extent $\Theta$ is adaptively calculated from the instantaneous boosted FAC grid as
\begin{equation} \label{eqn:Theta_calc}
    \Theta = \mathrm{max}\left[\tan^{-1}\left(\frac{v_{\perp}}{v_{\parallel}}\right)\right] \, .
\end{equation}

% This gives us
% \begin{equation} \label{eqn:frame_boost}
%     \Delta v_{\parallel} = \mathrm{max}\left\{ \frac{\sqrt{v_{\perp 1}^2 + v_{\perp 2}^2}}{\tan{\Theta}} + v_{\parallel} \right\}  + \mathrm{max}\{v_{\parallel}\} \, .
% \end{equation}

% When the magnetic field aligns perfectly with the radial direction of the instrument grid, this coordinate transformation is identical to the instrument's frame of reference. 

% Apart from the angular extent $\Theta$, generating a polar cap Slepian basis requires defining the maximum angular degree $L_{\mathrm{max}}$ of the wavelengths composing the basis functions. Since the boosted FAC ensures that the peak of the measured VDF is at the distance from the origin as in the instrument frame, we also choose the maximum wavenumber $L_{\mathrm{max}} = \pi / \Delta \phi$ where the Nyquist resolution $\Delta \phi$ depends on the instrument grid's angular resolution. For instruments with variable angular resolution in elevation and azimuth, the Nyquist resolution could be chosen as the larger of the resolutions in elevation and azimuth. For instance, in SPAN-Ai this would be $\sim 15^{\circ}$ leading to a maximum angular degree $L_{\mathrm{max}} = 12$. Generating Slepian basis functions using these choices of $\Theta$ and $L_{\rm{max}}$ ensures we recover the VDFs equivalent to instrument measurements at the maximum allowable resolution.

To create a polar cap Slepian basis, we need to define the angular extent ($\Theta$) and the maximum angular degree $L_{\mathrm{max}}$ of the basis function wavelengths. The boosted FAC ensures the peak of the measured VDF aligns with its position in the instrument frame. Thus, we set $L_{\mathrm{max}} = \pi / \Delta \phi$, with $\Delta \phi$ being the Nyquist resolution based on the instrument's angular resolution. If the angular resolutions are different for elevation and azimuth, we choose the coarser of the two. For instance, in SPAN-Ai, where the angular resolution in elevation is $15^{\circ}$ and in azimuth is $11.25^{\circ}$, the Nyquist resolution of approximately $15^{\circ}$ and consequently $L_{\mathrm{max}} = 12$. Selecting these parameters guarantees that the Slepian basis functions accurately capture VDFs at the highest possible resolution. Therefore, for the same instrument, while the Slepian polar cap extent $\Theta$ is adaptively adjusted based on the significant count grids, the angular resolution $L_{\mathrm{max}}$ is fixed by instrument design.

\subsubsection{Discretization of the gyrotropic VDF}

Since the magnetic field and solar wind speed can vary from measurement to measurement, rotating the ESA instrument grids into FACs may require a different mapping for every timestamp. Hence, we need to identify the gyroaxis for each independent measurement. For this, we have selected the 1D polar-cap basis functions to compose the 2.5D GDFs according to Eqn.~(\ref{eqn:f_gyro_1DSlep}), as this method is computationally more efficient.  
%Every timestamp has a different magnetic field, and therefore, each FAC grid distribution is different. To speed up the process of finding the gyroaxis, we have chosen to use the 1D polar-cap basis functions to compose the 2.5D GDF according to Eqn.~(\ref{eqn:f_gyro_1DSlep}). 
Fig.~\ref{fig:Fitting_Setup} shows the basis discretization of the gyrotropic distribution in FAC. For this timestamp, we have chosen $\boldsymbol{\hat{b}} = (0.981, -0.173, -0.091)$ and bulk velocity $\mathbf{u}_{\mathrm{bulk}} = (-433.51, 107.62, 32.55)$ [km/s]. Panel (A) shows the distribution of instrument grids in the boosted FAC, which have at least a single count measurement\footnote{\texttt{gdf} has the option to specify a \texttt{COUNT\_THRESHOLD} for filtering ESA grids in the reconstruction process}. For the particular case shown in Fig.~\ref{fig:Fitting_Setup}, we use a polar cap of angular extent $\Theta = 57.03^{\circ}$. The resultant $(v_{\parallel}, v_{\perp})$ along with the $57.03^{\circ}$ one-sided polar cap is shown in panel (A). Note that the grids are one-sided in $v_{\perp}$ by construction and the complete distribution, under gyrotropy, would imply a mirror reflection about the $v_{\perp} = 0$ (black vertical dashed line).

The domain is discretized as a function of velocity $v = \sqrt{v_{\parallel}^2 + v_{\perp}^2}$ and polar angle $\theta = \tan^{-1}\left(\frac{v_{\perp}}{v_{\parallel}}\right)$. To render localized support on different velocity shells, we use cubic B-splines anchored at knots computed for each timestamp based on the extent of the grids in the boosted FAC. The knots are placed logarithmically with a spacing of $1.2 \, \overline{\Delta \log_{10}(v)}$. Note that, using trial and error, we found that the slight adjustment factor of $1.2$ reduces collinearity between subsequent B-splines. This is especially important at low energies since the knots in $v$ are located in very close proximity. This results in nearly identical shape-functions (composition of the 2D basis using the 1D Slepian + 1D B-splines), which would make for an ill-posed inverse problem. Subsequently, this small adjustment factor of 1.2, therefore, makes the inversion better-conditioned while ensuring we do not undersample in $v$. The average log spacing between the ESA velocity shells is calculated as 
\begin{equation} \label{eqn:dlnv}
    \overline{\Delta \log_{10}(v)} = \frac{1}{N_{\rm{shells}}}\sum_{i = 0}^{N_{\rm{shells}}} \left[\log_{10}(v_{i+1}) - \log_{10}(v_i) \right] \, ,
\end{equation}
and we observed that for continuity in $v$, we want each B-spline to be sensitive to the grid points around the knot it's anchored on, as well as have partial sensitivity to grids located on two neighboring knots. The number of knots is determined using logarithmic spacing and the maximum extents of the grid points that have a non-zero count.
\begin{equation}
    N_{\rm{knots}} \simeq \frac{\log_{10}(v_{\rm{max}}) - \log_{10}(v_{\rm{min}})}{\overline{\Delta \log_{10}(v)}} \, ,
\end{equation}
where $N_{\rm{knots}}$ is rounded down to the closest integer and $(v_{\rm{min}}, v_{\rm{max}})$ are the minimum and maximum velocity magnitudes of grids with non-zero counts. Finally, the knot locations in $v$ are computed from binning $\log_{10}(v)$ into $N_{\rm{knots}}$ and finding the bin centers. The above prescription renders the first and the last points at the edge of the domain, resulting in zero B-spline support. To ensure that B-splines support the first and last points, we add two additional knots at $\log_{10}(v_{\rm{min}}) - 0.5\,\overline{\Delta \log_{10}(v)}$ and $\log_{10}(v_{\rm{max}}) + 0.5\,\overline{\Delta \log_{10}(v)}$. Finally, these knots are raised to the power of 10 and are used to generate the B-splines for each timestamp. Panel (B) of Fig.~\ref{fig:Fitting_Setup} shows an example set of B-splines for the particular test case. We have highlighted one of the B-splines peaking around $v \sim 500$ km/s in red. The arc of $(v_{\parallel}, v_{\perp})$ space is also shaded in red in panel (A). The cloud of grids that are spanned by this B-spline (using 0.1 in panel (B) as a threshold) is marked in large black dots in panel (A).

In panel (C) of Fig.~\ref{fig:Fitting_Setup}, we show the first $N^{\mathrm{2D}}_{\mathrm{polcap}} = 4$ Slepian functions that are optimally confined inside a $\Theta = 57.03^{\circ}$ polar cap with a maximum wavenumber $L_{\rm{max}} = 12$. These functions capture the variation of the distribution across the grid points as a function of polar angle for each B-spline velocity shell. The $\Theta = 57.03^{\circ}$ is demarcated by a vertical red dashed line. The angular location of the grids spanned within the red arc in panel (A) is overplotted on each Slepian function.

\subsubsection{Estimation of the gyroaxis}\label{sec:gyroaxis_estimation}
\begin{figure}
    \centering
    \includegraphics[width=\linewidth]{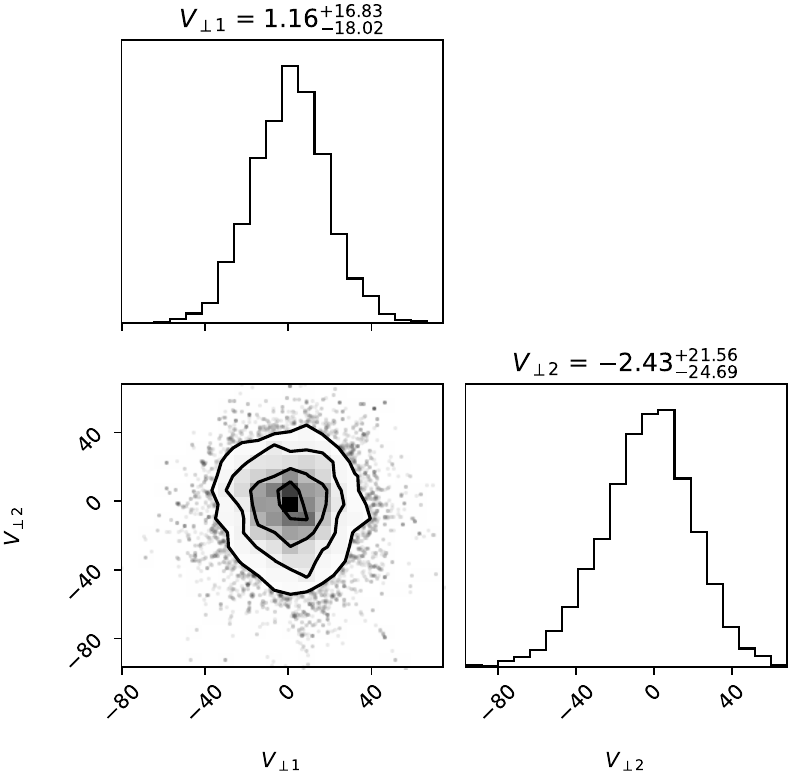}
    \caption{MCMC corner plots \citep{corner} of the Bayesian gyroaxis optimization using 8 walkers and 2000 steps each. The final solution from the \texttt{scipy} minimization $\widetilde{\boldsymbol{U}}$ was used as the initial guess about which walkers were randomly initialized. The $0.5$ quantile is indicated along with the 0.14 ($\approx -1\sigma$) and 0.86 ($\approx +1\sigma$) quantile error bars.}
    \label{fig:mcmc}
\end{figure}
Using the 2.5D modelling framework described above, we can iteratively find the gyroaxis location for which the reconstructed distribution best matches the ESA measurements. Since we know the direction of the magnetic field, there are only two degrees of freedom for the location of the gyroaxis, which are in the plane perpendicular to $\hat{\boldsymbol{b}}$. The perpendicular directions $(\hat{\boldsymbol{p}}, \, \hat{\boldsymbol{q}})$ are calculated as
\begin{equation}
    \hat{\boldsymbol{p}} = \frac{\hat{\boldsymbol{b}} \times \hat{\boldsymbol{x}}}{|\hat{\boldsymbol{b}} \times \hat{\boldsymbol{x}}|}, \qquad \hat{\boldsymbol{q}} = \hat{\boldsymbol{b}} \times \hat{\boldsymbol{p}} \, ,
\end{equation}
where, $\hat{\boldsymbol{x}}$ is the unit vector in the instrument frame along the $x$-direction.
During intervals where $\hat{\boldsymbol{b}}$ is very closely aligned with the $x$-direction, we define $\hat{\boldsymbol{p}} = \hat{\boldsymbol{b}} \times \hat{\boldsymbol{y}} \big/ |\hat{\boldsymbol{b}} \times \hat{\boldsymbol{y}}|$. At every iteration of the optimization problem, our model requires a bulk velocity to perform FAC coordinate readjustment before fitting Eqn.~(\ref{eqn:f_gyro_1DSlep}). This bulk velocity is defined as
\begin{equation} \label{eq:bulk_velocity}
    \boldsymbol{U} = \boldsymbol{U}_{\rm{init}} + V_{\perp 1} \, \boldsymbol{\hat{p}} + V_{\perp 2} \, \boldsymbol{\hat{q}} \, .
\end{equation}
Therefore, a complete reconstruction for each iteration involves inferring $V_{\perp 1}, V_{\perp 2}$. As a first step, we perform minimization by implementing the L-BFGS (Limited-memory Broyden–Fletcher–Goldfarb–Shanno) algorithm using the \texttt{minimize} module of \texttt{scipy.optimize} Python package. This provides us a first estimate of the \textit{gyro-centroid} $(V_{\perp 1}, V_{\perp 2})$. Hereafter, we will refer to this \texttt{scipy} solution as $\widetilde{\boldsymbol{U}}$.

The \texttt{gdf} package has the option to further refine this L-BFGS solution via an MCMC optimization flag which uses the \texttt{emcee} Python package \citep{emcee}. In this case, for a starting guess of bulk velocity, we use $\boldsymbol{U}_{\rm{init}} = \widetilde{\boldsymbol{U}}$. The MCMC walkers are initialized randomly about this point. We use uninformed, flat priors during the MCMC iterations. The final converged solution $\hat{\boldsymbol{U}}$ from MCMC is obtained by plugging in the 0.5 quantile values of the posterior distributions for $V_{\perp 1}$ and $V_{\perp 2}$ in Eqn.~(\ref{eq:bulk_velocity}). For these optimum parameters, we can compute the model covariance matrix from the MCMC chains. This can then be used to calculate the component uncertainties along $U_x, U_y, U_z$ as below
\begin{equation}
\sigma_i^2 =
\begin{bmatrix}
p_i & q_i
\end{bmatrix}
\begin{bmatrix}
\mathrm{Var}(V_{\perp 1}) & \mathrm{Cov}(V_{\perp 1}, V_{\perp 2}) \\
\mathrm{Cov}(V_{\perp 1}, V_{\perp 2}) & \mathrm{Var}(V_{\perp 2})
\end{bmatrix}
\begin{bmatrix}
p_i \\
q_i
\end{bmatrix} \, ,
\end{equation}
where, $i \in (x, y, z)$ and $p_i, q_i$ denotes the projections of the unit vectors along axis $i$.

To summarize this subsection, the entire workflow for fitting the gyro-centroid is outlined in Fig.~\ref{fig:Flowdiagram}, and the steps are restated as follows:
\begin{enumerate}[noitemsep]
    \item Load in the instrument frame grids and the associated VDF 
    % data\footnote{\href{http://sweap.cfa.harvard.edu/pub/data/sci/sweap/spi/L2/spi_sf00/}{http://sweap.cfa.harvard.edu/pub/data/sci/sweap/spi/L2/spi\_sf00/}} 
    along with the count information.
    \item Using the instrument grid integrated bulk velocity moment and the averaged magnetic field over the collection window, convert the instrument frame grids to FAC. Perform the required boost $\Delta v_{\parallel}$ from Eqn.~(\ref{eqn:frame_boost}) and compute the corresponding polar cap of $\Theta$ calculated according to Eqn.~(\ref{eqn:Theta_calc}). This gives us the $(v_{\parallel}, v_{\perp})$ grid to perform the 2.5D fitting.
    \item Generate B-splines and Slepian functions inside the polar cap. For the gyro-centroid finder at each timestamp, the Slepian basis coefficients are computed only once for a given $\Theta$ before the minimization. They are used as a generative function to obtain the functional values for each grid combination. The knots for B-splines are generated at every iteration to accommodate the varying grid density in $v$.
    \item A first estimate of the gyro-centroid is obtained after performing a minimization using L-BFGS. A good measure of the initial guess for the minimization algorithm is the instrument-integrated velocity moment.
    \item An optional refinement of the gyro-centroid is available where an MCMC-based posterior is obtained in $(V_{\perp 1},\, V_{\perp 2})$ along with accompanying error bars. These are then projected to obtain $\sigma_x, \sigma_y$ and $\sigma_z$ error estimates of the bulk velocity.   
\end{enumerate}
At this point, we are equipped with a robust estimate for the gyro-centroid where we anchor the magnetic field and build our gyrotropic distribution about it. The next section describes the final VDF inversion using one of the three currently available reconstruction modules in \texttt{gdf}.

\subsection{Super-resolution} \label{sec:super-resolution}
In the above section, we present procedures that are defined generally for the entire GDF reconstruction process. In this subsequent section, we focus on three methods within the \texttt{gdf} package that each generate a gyrotropic distribution. The three modules of super-resolving in \texttt{gdf} are (A) polar cap: using 1D polar-cap Slepians + 1D B-splines, (B) Cartesian: using 2D Cartesian Slepian functions generated inside a defined convex hull containing the statistically significant grid points, and (C) hybrid: a combination of the polar cap and Cartesian methods to better regularize the inversion. 

It is important to note that we perform all Slepian fits to the data on a log scale. This is primarily because (A) if we discretize the linear scale VDF in terms of Slepian functions while still requiring a linear inversion (independent of model parameter initializations), positivity would not be guaranteed, and (B) the VDF usually spans multiple orders of magnitude resulting in the linear scale fits almost entirely bring sensitive to the core (higher amplitude population) and insensitive to the beam (lower amplitude population). However, fitting in log-scale means that (log-scale) misfits at the core of the same magnitude is regarded on the same footing as (log-scale) misfits at the beam. This can result in very different bulk moments even if the log-scale fits produce excellent goodness-of-fit statistics. In order to overcome this issue, we perform a first step where we fit the data in linear scale to a single anisotropic Maxwellian $\mathcal{M}(n^0, v_{\parallel}^{0}, w_{\parallel}^0, w_{\perp}^0)$. This fit is almost entirely sensitive to the high intensity cap of the core population. Thereafter, we add a set of fiducial points $(v_{\perp}^{\mathrm{fid}}, v_{\parallel}^{\mathrm{fid}})$ on a line between $(0, v_{\parallel}^{0} - w_{\parallel}^0/2)$ and $(0, v_{\parallel}^{0})$. For all of the super-resolution methods below, we append the fiducial values $\log_{10}{\mathcal{M}(v_{\perp}^{\mathrm{fid}}, v_{\parallel}^{\mathrm{fid}})}$ to the log-scaled VDF measurements used for fitting the Slepian basis functions. The set of these log-scaled ESA measurements are referred to as $\mathbf{d_{\mathrm{ESA}}}$ in the following subsections. This ensures that we preserve the kinetic structure in log-scale while not compromising moments.

The addition of fiducial points is one of several options that can be employed to address the log-to-linear conversion issue mentioned above. Apart from the approach used in this study, one could (A) place a single point at (0,0) with a higher weight than the other ESA grid points, and (B) using complementary measurements such as FOV integrated fluxes from SPC, a part of the SWEAP suite along with SPAN-i. These are useful in cases where there is reason to believe that the peak of the distribution could deviate significantly from being a Maxwellian (such as in a flat-top distribution). Note that the closer the measured ESA grids are to $v_{\perp} = 0$ near the core, the less sensitive is the final reconstruction to any of these choices.

\subsubsection{Method A: \textsc{polcap}} \label{sec:polcap}

\begin{figure*}
    \centering
    \includegraphics[width=\linewidth]{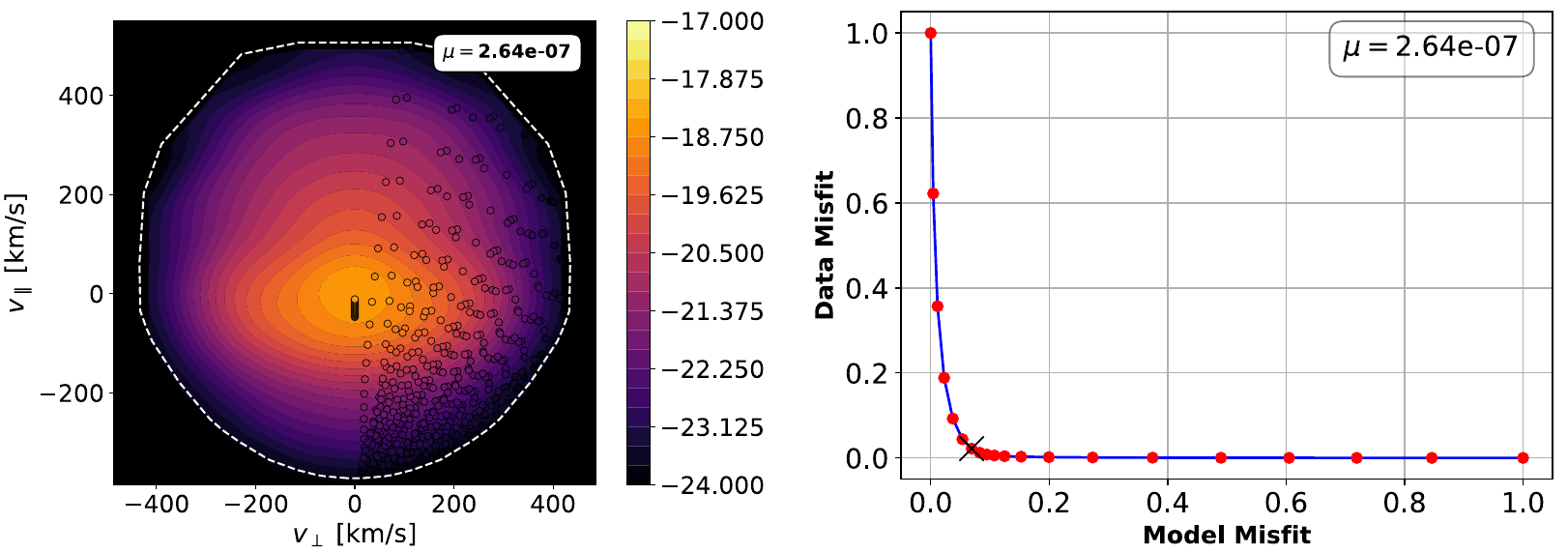}
    \caption{Super-resolution of a synthetic VDF using for $\boldsymbol{\hat{b}} = 0.987 \, \boldsymbol{\hat{x}} -0.154 \, \boldsymbol{\hat{y}} -0.029 \, \boldsymbol{\hat{z}}$ and $\mathbf{u}_{\mathrm{bulk}}\mathrm{[km/s]} = -433.505 \, \boldsymbol{\hat{x}} +  107.615 \, \boldsymbol{\hat{y}} + 32.554 \, \boldsymbol{\hat{z}}$ for our synthetic VDF described in Sec.~\ref{sec:synthetic_demo}. The \textit{left} panel shows the super-resolved GDF as the background colormap. The synthetic data is overplotted as scattered points on an irregular grid as obtained from the boosted FAC. According to the calculations in Eqns.~(\ref{eqn:frame_boost})-(\ref{eqn:Theta_calc}), $\Delta v_{\parallel} = 463.37$ km/s, $\Theta = 57.03^{\circ}$ and $N^{\mathrm{2D}}_{\mathrm{polcap}} = 4$ with 21 knots placed logarithmically according to Eqn.~(\ref{eqn:dlnv}) across all the FAC grids. The \textit{white-dashed} line marks the convex hull encompassing all the grids with significant counts. The \textit{right} panel shows the trade-off plot between data-misfit and model-misfit. The knee of this L-curve, indicated by the \textit{black} `x', is used to find the optimal regularization constant $\mu$ to be used in Eqn.~(\ref{eqn:polcap_inv}) to obtain the model coefficients $\mathbf{M_A}$. In this case, the B-spline regularization is found to be $\mu = 2.64 \times 10^{-7}$ from the knee of the trade-off curve.}
    \label{fig:polcap_method}
\end{figure*}

This method is the same as that outlined in Section~\ref{sec:gyro-centroid-finder} to find the gyro-centroid. Therefore, we use Eqn.~(\ref{eqn:f_gyro_1DSlep}) and the ESA data to find the coefficients $c^{i,\alpha}$. After performing this linear inversion, we use these $c^{i,\alpha}$ to super-resolve the gyrotropic distribution at a densely sampled grid of $(v, \theta)$. We collapse the knots and Slepian dimensions into a 1D array $M_{\mathrm{A}}^k = c^{i, \alpha}$ to obtain a simpler labeling convention for these model coefficients for method A. If $i \in (0, ..., N_{\mathrm{splines}})$ and $\alpha \in (0, ..., N_{\mathrm{Slepians}})$, then $k$ is a composite index such that $k = \alpha \, + \, N_{\mathrm{Slepians}} \, (i - 1)$ and $p \in (0, ..., N_{\mathrm{knots}} \times N_{\mathrm{Slepians}})$. Since $M_{\mathrm{A}}^p$ is an element belonging to the matrix $\mathbf{M}^{\mathrm{A}}$, we can write the expression for super-resolution as
\begin{equation}
    \log_{10}{f_{\mathrm{supres}}^{\mathrm{A}}(v_{\mathrm{supres}}, \theta_{\mathrm{supres}})} = \mathbf{G_A}(v_{\mathrm{supres}}, \theta_{\mathrm{supres}}) \cdot \mathbf{M_A} \, ,
\end{equation}
where, $G_{\mathrm{A};i,\alpha} = \beta_i(v_{\mathrm{supres}}) \, g^{\mathrm{polcap}}_{\alpha}(\theta_{\mathrm{supres}})$ and we perform a similar composite index transformation $(i, \alpha) \rightarrow k$ to get $G_{A;k}(v_{\mathrm{supres}}, \theta_{\mathrm{supres}})$ which are elements of the matrix $\mathbf{G_A}(v_{\mathrm{supres}}, \theta_{\mathrm{supres}})$.

It is important to note that depending on the direction of the magnetic field vector $\hat{\boldsymbol{b}}$ and the fraction of VDF inside the instrument's FOV, the linear inverse problem of obtaining the model coefficients $\mathbf{M_A}$ could be poorly conditioned in $v$. Therefore, we use a second-derivative smoothness regularization which introduces the matrix $\mathbf{D}$ of shape $(k \times k)$. More details on the derivation of this regularization matrix can be found in Appendix~\ref{sec:B-spline-Reg-matrix}. Including the second derivative penalty, the cost function for the polar-cap model can be expressed as
\begin{eqnarray}
    \chi^2_{\mathrm{polcap}} &=& \left(\mathbf{d}_{\mathrm{ESA}} - \mathbf{G_A} \cdot \mathbf{M_A} \right)^T \cdot \left(\mathbf{d}_{\mathrm{ESA}} - \mathbf{G_A} \cdot \mathbf{M_A} \right) \nonumber \\
    &+&   \mathbf{M_A}^T \cdot \mathbf{D} \cdot \mathbf{M_A} \, ,
\end{eqnarray}
where, $\mu$ is the regularization constant determined using the L-curve method from the data-misfit vs. model smoothness trade-off curve. Minimizing the cost function with respect to model coefficients $\mathbf{M_A}$, we obtain a least squares regularized linear inverse problem
\begin{equation} \label{eqn:polcap_inv}
    \mathbf{M_A} = \left(\mathbf{G_A}^{\mathrm{T}} \cdot \mathbf{G_A} + \mu \, \mathbf{D}\right)^{-1} \cdot \mathbf{G_A}^{\mathrm{T}} \cdot \mathbf{d_{\mathrm{ESA}}} \, .
\end{equation}
Figure~\ref{fig:polcap_method} demonstrates the reconstruction of a synthetic VDF on an ESA grid. The background colormap in the \textit{left} panel shows the super-resolution. The synthetic ESA data are over-plotted in the boosted FAC frame. The \textit{white-dashed} boundary denotes the convex hull enclosing all the grids with counts higher than the count threshold (in this case we use count threshold as 1). The polar-cap basis functions (combination of 1D Slepian functions in the angular space localized on 1D B-splines in the velocity space) are shown in Fig.~\ref{fig:Fitting_Setup}. The final choice of $\mu \simeq 2.64 \times 10^{-7}$ is obtained from the knee of the trade-off curve as shown in the \textit{right} panel of Fig.~\ref{fig:polcap_method}.

\subsubsection{Method B: \textsc{cartesian}} \label{sec:cartesian}

\begin{figure*}
    \centering
    \includegraphics[width=1\linewidth]{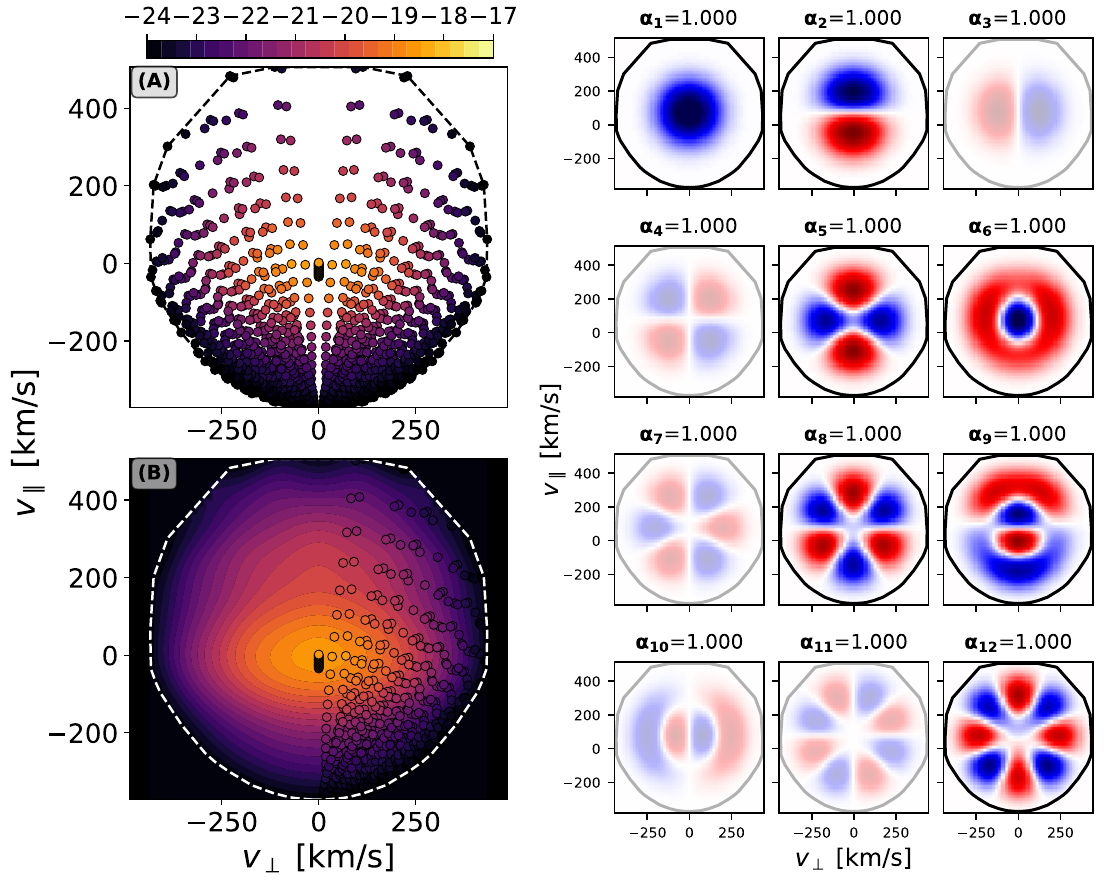}
    \caption{GDF reconstruction using 2D Cartesian Slepian basis functions optimally concentrated within the convex hull formed by the outermost bounds of the grids containing significant counts. The convex hull is shown using a \textit{dashed} boundary circumscribing the scattered grid points. Panel (A) shows the distribution of grids in boosted FAC coordinates, for the same synthetic ESA measurement and $\boldsymbol{\hat{b}}$ as in panel (A) of Fig.~\ref{fig:Fitting_Setup}. Panel (B) shows the results of the super-resolved reconstruction using 2D Cartesian Slepian bases. The discrete synthetic ESA measurements from panel (A) are over-plotted on the right-half of reconstruction in panel (B). As in Fig.~\ref{fig:polcap_method}, the left half is not over-plotted with the measurements for visual clarity. The collage of figures on the \textit{right} panel, labelled with $\alpha_1 - \alpha_{12}$, shows a few Cartesian Slepian basis functions inside the convex hull for synthetic ESA measurements above the count threshold. The basis functions are generated at a maximum allowable ``phase-space" wavenumber $k_{\mathrm{max}} = 0.0319 \, \mathrm{rad \, s/km}$ which corresponds to a ``phase-space" wavelength of $\sim 98 \, \mathrm{km/s}$. Using Eqn.~(\ref{eqn:N2D_cartesian}), we have $N^{\mathrm{2D}}_{\mathrm{cart}} \equiv 49$. We show the 12 most optimally concentrated Slepian bases. The color-scale of the basis functions are in arbitrary units. Regions colored in \textit{red} and \textit{blue} have opposite signs and \textit{white} indicates nodes passing through zero. The eigenvalue (interpretable as the energy confined inside the domain of concentration) for each basis function is denoted as $\alpha_i$. Since our problem is symmetric about the $v_{\perp}$ axis, we only use the even basis functions and are plotted using high opacity. The odd basis functions are faded out.}
    \label{fig:Sleprec2D}
\end{figure*}

Having found the gyroaxis as in Section~\ref{sec:gyro-centroid-finder}, we can alternatively use 2D Cartesian Slepian bases localized within a compact domain in $\mathbb{R}^2$ space for reconstructing the GDF. We refer the reader to Section~4 in the original work of \cite{Simons&Wang2011} for further details about the mathematical formulation of these 2D Slepian functions. As in the case of the polar cap Slepian, the 2D Cartesian Slepian basis are also evaluated numerically as the eigenfunctions of a Fredholm integral eigenvalue equation. The 2D Cartesian Slepian basis depends on two input factors --- the domain of concentraion in $\mathcal{R}^2$ and the resolvable scale of the basis functions (proportional to the number of basis functions concentrated inside $\mathcal{R}^2$). The corresponding \textsc{Matlab} software used in this paper to generate the 2D Cartesian Slepian bases can be found in \cite{slepian_alpha} and \cite{slepian_foxtrot}. In particular, we used the \textsc{Matlab} function \texttt{localization2D.m} in the \textsc{slepian\_foxtrot} repository to generate the 2D Cartesian Slepian functions. In our case, $\mathbb{R}^2$ is in $(v_{\parallel},\, v_{\perp})$ FAC velocity phase space. Therefore, we can express the 2D gyrotropic distribution in the basis of 2D Cartesian Slepians as
\begin{equation}
    \log_{10}{f_{\mathrm{gyro}}(v_{\parallel}, v_{\perp})} = \sum_{\alpha=0}^{N^{\mathrm{2D}}_{\mathrm{cart}}} c^\alpha \, g_{\alpha}^{\mathrm{cart}}(v_{\parallel}, v_{\perp}) \, .
\end{equation}
We choose to denote the model coefficient vector of $c_{\alpha}$ as $\mathbf{M_B}$ of length $N^{\mathrm{2D}}_{\mathrm{cart}}$ and the matrix of Slepian functions $g_{\alpha}^{\mathrm{cart}}(v_{\parallel}, v_{\perp})$ as $\mathbf{G_B}$ of shape $(N_{\mathrm{ESA}} \times N^{\mathrm{2D}}_{\mathrm{cart}})$. Here, $N_{\mathrm{ESA}}$ is the total number of data points in $\mathbf{d_{\mathrm{ESA}}}$. The final inverse problem using just the Cartesian Slepian basis functions then takes the form
\begin{equation}
    \mathbf{M_B} = \left(\mathbf{G_B}^T \cdot \mathbf{G_B} \right)^{-1} \cdot \mathbf{G_B}^T \cdot \mathbf{d}_{\mathrm{ESA}} \, .
\end{equation}

Figure~\ref{fig:Sleprec2D} represents the reconstruction process using the 2D Slepian functions. The grids after converting to the FAC coordinates is plotted in Fig.~\ref{fig:Sleprec2D}(A). The measurements are colored by the synthetic VDF. Since we use a count threshold of 1, meaning we only use the points which have a finite count for our Slepian reconstruction, we can construct a convex hull $\mathcal{R}^2$ within which we generate the Slepian functions. The supported data (the regions of the instrument grid with significant counts) define the convex hull of the distribution. In field-aligned coordinates, the points at the outer edges of the measured domain form the vertices of this hull. Restricting our reconstruction to within the convex hull ensures that the linear inversion is confined to regions supported by data only. In Fig.~\ref{fig:Sleprec2D}(A) this is marked by the \textit{black dashed} line. Note that this convex hull $\mathcal{R}^2$ is specific to this timestamp of observation and changes for each time by virtue of the grids changing in FAC. The colorbar on top shows the exponent of 10 for each of the plotted levels of the VDF.

We then perform a reconstruction of the full distribution and super-resolve it to a fine, regular grid in $(v_{\parallel}, \, v_{\perp})$. This is shown in Fig.~\ref{fig:Sleprec2D}(B). We use the same colorbar as in panel (A). The synthetic ESA measurements are over-plotted only on the \textit{right} half of this figure for a visual goodness-of-fit assessment. In Fig.~\ref{fig:Sleprec2D}(B), the \textit{white dashed} line shows the convex hull $\mathcal{R}^2$.

The collage of figures on the right half of Fig.~\ref{fig:Sleprec2D} shows all the 2D Slepian basis functions optimally concentrated within $\mathcal{R}^2$ with a maximum horizontal wavenumber $k_{\mathrm{max}} = 0.0319 \, \mathrm{rad \, s/km}$ in velocity phase-space. This is computed from estimating the minimum resolvable wavelength in velocity phase-space of $\sim 98 \mathrm{km/s}$. \cite{Simons&Wang2011} show that the planar Shannon number, which we refer to as $N^{\mathrm{2D}}_{\mathrm{cart}}$ in this study, can be expressed as
\begin{equation} \label{eqn:N2D_cartesian}
    N^{\mathrm{2D}}_{\mathrm{cart}} = \frac{k_{\mathrm{max}}^2 \, A}{4\pi} \, ,
\end{equation}
where, $A$ is the area inside the convex hull $\mathcal{R}^2$. Prescribing the planar Shannon number and the convex hull $\mathcal{R}^2$, we can compute the 2D Cartesian Slepian bases optimally concentrated inside the convex hull. Eqn.~(38a) of \cite{Simons&Wang2011} present the construction of Slepian functions as an eigenvalue problem where each Slepian function is an eigenfunction $g_i$ and has an associated eigenvalue. We denote this eigenvalue as $\alpha_i$, which indicates the fraction of power of $g_i(v_{\parallel,} \, v_{\perp})$ within $\mathcal{R}^2$ as compared to $\mathbb{R}^2$. Note that the GDF is symmetric about the $v_{\perp}=0$ line by construction. Therefore, we only need the basis functions which are symmetric about $v_{\perp}$. In the figure, we have reduced the opacity of the odd basis functions in order to show this, emphasizing only the even functions that contribute to our GDF reconstruction. 

\subsubsection{Method C: \textsc{hybrid}} \label{sec:hybrid}

The examples shown in Sections.~\ref{sec:polcap} \& \ref{sec:cartesian} present the potential for reconstructions using polar-cap Slepians and Cartesian Slepians. From these examples, we know that there are instances where these reconstructions present satisfactory results. 

However, when scrutinizing robustness on a case-by-case basis, we observed that both of these methods have characteristic drawbacks. In cases where the angular separation between the anodes containing the beam is less than the Nyquist resolution (given by $L_{\mathrm{max}}$), the polar cap method successfully captures the beam. However, this reconstruction shows small wiggles around the core. This is by virtue of the dense knot spacing of B-splines around the core (with higher density of measurements) which allows more flexibility to fit the data points at the expense of introducing such artifacts of overfitting. The ESA operations concept is generally to sample high energy portions of the phase space more sparsely than low energy. Therefore, using the polar cap method often results in the distribution having fine variations at lower energies as compared to high energies. However, there is no reason to believe this is a real feature. 
% Moreover, the polar cap method involves using B-splines which have strictly zero values under a given knot. This results in a characteristic inward bend which shows up in certain cases of grid configuration. 

Going forward, we make no apriori assumptions and impose no bias on VDF granularity as a function of energy. Unlike the polar cap method, the 2D Cartesian method does not have an energy dependent granularity since it uses basis functions with equal maximal angular degree in all of $\mathbb{R}^2$ phase space. Consequently, when the sampling is sparse at higher energies, the Cartesian basis functions can be under-constrained there. This is why the Cartesian method sometimes fails to capture the beam as effectively as the polar-cap method. 

The purpose of the hybrid method is to synthesize a combination of Methods A \& B that circumvents their respective weaknesses. For the test cases presented in the previous subsections (and later in Fig.~\ref{fig:grazing-angle-demo}), we note that (A) the small wiggles seen around the core in the polar-cap method are below the resolution limit of the 2D Cartesian wavelengths, and (B) the beam artifact introduced in the 2D Cartesian method is below the angular resolution of the 1D polar-cap Slepian basis. It is apparent from this that the limitation of one method falls in the null-space of the other. Motivated by that observation, we add a similarity term to the hybrid cost function. The hybrid algorithm performs simultaneous inversions for both $\mathbf{M_A}$ and $\mathbf{M_B}$ such that the super-resolutions $\mathbf{G_A}^{\mathrm{supres}} \cdot \mathbf{M_A}$ and $\mathbf{G_B}^{\mathrm{supres}} \cdot \mathbf{M_B}$ evaluated on the same super-resolution grid are sufficiently similar. The degree of similarity is imposed by the similarity index $\lambda$. For $\lambda = 0$, the Method A and Method B inversions are independent. For $\lambda \gg 1$, their super-resolution outputs must be identical. The total cost function for such an inverse problem takes the form
\begin{equation}\label{eqn:hybrid_cost}
    \chi^2_{\mathrm{total}} = \chi^2_\mathrm{A} + \chi^2_\mathrm{B} + \lambda \, \chi^2_{\mathrm{sim}}
\end{equation}
where,
\begin{eqnarray*}
    \chi^2_\mathrm{A} &=& \left(\mathbf{d}_{\mathrm{ESA}} - \mathbf{G_A} \cdot \mathbf{M_A} \right)^T \cdot \left(\mathbf{d}_{\mathrm{ESA}} - \mathbf{G_A} \cdot \mathbf{M_A} \right) \, , \\
    \chi^2_\mathrm{B} &=& \left(\mathbf{d}_{\mathrm{ESA}} - \mathbf{G_B} \cdot \mathbf{M_B} \right)^T \cdot \left(\mathbf{d}_{\mathrm{ESA}} - \mathbf{G_B} \cdot \mathbf{M_B} \right) \, ,\\
    % \chi^2_{\mathrm{reg}} &=& \mathbf{M_A}^T \cdot \mathbf{D} \cdot \mathbf{M_A} \, ,\\
    \chi^2_{\mathrm{sim}} &=& \left(\mathbf{G_A}^{\mathrm{supres}}  \cdot \mathbf{M_A} - \mathbf{G_B}^{\mathrm{supres}}  \cdot \mathbf{M_B} \right)^T \cdot \\
    &&\left(\mathbf{G_A}^{\mathrm{supres}}  \cdot \mathbf{M_A} - \mathbf{G_B}^{\mathrm{supres}}  \cdot \mathbf{M_B} \right)
\end{eqnarray*}
Minimizing $\chi^2_{\mathrm{total}}$ with respect to $\mathbf{M_A}$ and $\mathbf{M_B}$, the final hybrid inversion, in matrix form, looks like
\begin{equation} \label{eqn:hybrid_inversion}
    \begin{bmatrix}
    \mathbf{M_A} \\
    \mathbf{M_B}
    \end{bmatrix} 
    = \left(\mathbf{G_C}^T \cdot \mathbf{G_C} \right)^{-1} \cdot \mathbf{G_C}^T \cdot \mathbf{d_C} \, ,
\end{equation}
where the augmented Slepian matrices $\mathbf{G_C}$ and the augmented data matrix $\mathbf{d_C}$ can be expressed as
\begin{equation*}
\mathbf{G_C} = 
\begin{bmatrix}
\mathbf{G_A} & \mathbf{0} \\
\mathbf{0} & \mathbf{G_B} \\
\sqrt{\lambda} \mathbf{G_A}^{\mathrm{supres}} & -\sqrt{\lambda} \mathbf{G_B}^{\mathrm{supres}}
% \sqrt{\mu} \mathbf{D} & \mathbf{0}
\end{bmatrix},
\,
\mathbf{d_C} = 
\begin{bmatrix}
\mathbf{d}_{\mathrm{ESA}} \\
\mathbf{d}_{\mathrm{ESA}} \\
\mathbf{0}
% \mathbf{0}
\end{bmatrix} \, .
\end{equation*}

We find the $\lambda$ that produces the optimal balance between similarity across the polar cap and Cartesian reconstructions and the respective data misfits using the knee of the trade-off curve between ($\chi^2_\mathrm{A} + \chi^2_\mathrm{B}$) and $\chi^2_{\mathrm{sim}}$. 

\begin{figure*}
    \centering
    \includegraphics[width=\linewidth]{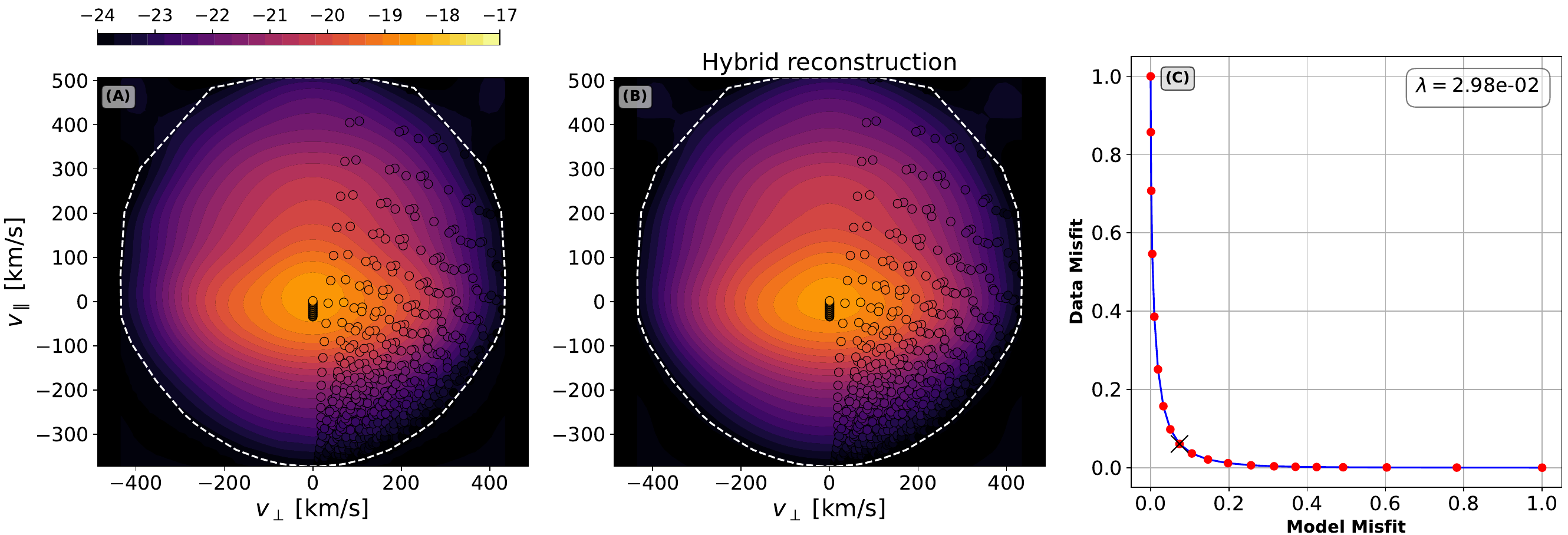}
    \caption{Super-resolution using the hybrid method outlined in Section~\ref{fig:hybrid_method} for the same synthetic ESA measurement and $\boldsymbol{\hat{b}}$ as in panel (A) of Fig.~\ref{fig:Fitting_Setup}. Eqn.~(\ref{eqn:hybrid_inversion}) is used to jointly solve the polar cap model parameters $\mathbf{M_A}$ and Cartesian model parameters $\mathbf{M_B}$ by inducing an adaptive similarity index using a L-curve method. Panel (A) shows the reconstruction using $\mathbf{G_A}^{\mathrm{supres}}(v_{\parallel}, v_{\perp})\cdot \mathbf{M_A}$ and Panel (B) shows the reconstruction using $\mathbf{G_B}^{\mathrm{supres}}(v_{\parallel}, v_{\perp})\cdot \mathbf{M_B}$. The similarity index $\lambda$ in Eqn.~(\ref{eqn:hybrid_cost}) is calculated by locating the knee of the data-misfit vs. model-misfit trade-off curve shown in Panel (C). The located knee at $\lambda = 2.98 \times 10^{-2}$ is marked by a black `x'. The \textit{white-dashed} line marks the convex hull within which the GDF is localized. The scatter points overlaid on the reconstructed GDF denotes the synthetic data in the FAC frame.}
    \label{fig:hybrid_method}
\end{figure*}

Figure~(\ref{fig:hybrid_method})(A) \& (B), respectively, shows the polar-cap and Cartesian reconstructions after imposing this similarity index. Panel (C) shows the trade-off curve and marks the located knee with a `x' at the optimal $\lambda = 2.98\times 10^{-2}$. In the current version of \texttt{gdf}, the similarity imposed Cartesian super-resolution is the final hybrid product.

\subsection{GDF moment calculations}\label{sec:gdf_moms}

After obtaining the super-resolved GDF in the boosted FAC using one of the three methods in Section~\ref{sec:super-resolution}, we evaluate the distribution on a uniform rectilinear grid, allowing for simple velocity moment calculations using a Riemann sum. At this point, the GDF is evaluated in the shifted plasma frame such that $v_{\parallel} \rightarrow v_{\parallel} - \Delta v_{\parallel}$. The gyro-centroid method fixes the $v_{\perp,1}$ and $v_{\perp,2}$ along the gyroaxis, held at $\tilde{v}_{\parallel} = \tilde{\bm{U}} \cdot \hat{\bm{b}}$. The moments are evaluated as
\begin{eqnarray}
    n &=& 2\pi \sum_{i,j} v_{\perp}^j \, f(v_{\parallel}^i, \, v_{\perp}^j) \, \mathrm{d}v_{\parallel}^i \, \mathrm{d}v_{\perp}^j \, ,\\
    v'_{\parallel} &=& 2\pi \sum_{i,j} v_{\parallel}^i \, v_{\perp}^j \, f(v_{\parallel}^i, \, v_{\perp}^j) \, \mathrm{d}v_{\parallel}^i \, \mathrm{d}v_{\perp}^j \, ,\\
    \frac{k_b T_{\perp}}{m_p} &=& 2\pi \sum_{i,j} \, (v_{\perp}^j)^{3} \, f(v_{\parallel}^i, \, v_{\perp}^j) \, \mathrm{d}v_{\parallel}^i \, \mathrm{d}v_{\perp}^j \, ,\\
\end{eqnarray}
and
\begin{eqnarray}
    \frac{k_b T_{\parallel}}{m_p} &=& 2\pi \sum_{i,j} (v_{\parallel}^i - v'_{\parallel})^{2} \, v_{\perp}^j \, f(v_{\parallel}^i, \, v_{\perp}^j) \, \mathrm{d}v_{\parallel}^i \, \mathrm{d}v_{\perp}^j \, , \nonumber \\
\end{eqnarray}
where $v'_{\parallel}$ is the moment evaluated in the boosted frame. Given that the distribution should be in the \textit{plasma} frame, the velocity moments for $v_{\parallel}$ and $v_{\perp}$ should evaluate to zero. Using this fact, we can define $\Delta v_{\parallel} = v'_{\parallel} - v_{\mathrm{shift}}$. If the value is nonzero, we rotate from FAC to the instrument frame to provide a final correction to the spacecraft frame moments.   
 
\section{Synthetic example of GDF reconstruction} \label{sec:synthetic_demo}
To demonstrate the gyrotropic SBR method and its robustness to partial VDF coverage, we reconstruct a model distribution expressed on a synthetic ESA instrument grid. This is inspired by the FOV limitation of SPAN-Ai. For simplicity, we choose a bi-Maxwellian approximation
expressed in field-aligned coordinates $(v_\parallel, v_\perp)$,  
\begin{equation}
    \mathcal{M}(v_{\parallel}, v_{\perp}) = \frac{n}{\pi^{3/2} w_\perp^{2} w_\parallel} \exp{\bigg[\bigg(\frac{-(v_\parallel - u_\parallel)^2}{w_\parallel^2}\bigg)-\bigg(\frac{v_\perp^2}{w_\perp^2}\bigg)\bigg]},
    \label{eq:biMax}
\end{equation}
where $n$ is the number density, $u_\parallel$ is the drift speed along the magnetic field, and $w_\parallel$ and $w_\perp$ are the thermal speeds parallel and perpendicular to the magnetic field, respectively. We include two bi-Maxwellian populations in the model-- a proton core and beam\footnote{Note the values are drawn from bi-Maxwellian fits to SPAN-Ai data for a solar wind stream at $\sim0.06~\textrm{au}$, with a proton plasma beta $\beta_p < 0.07$, and solar wind speed of $v_{sw} \sim 380~\textrm{km/s}$, following the methods of \citep{Woodham_2021, Laker2024}}. %{\color{red}with values drawn from bi-Maxwellian fits to Parker Solar Probe observations \cite{Jaye_Verniero_202?}}. 
For each of our synthetic tests, we define the core population with $n_{\mathrm{core}} = 1466.9 \,\rm{cm}^{-3}$, $w_{\parallel, \mathrm{core}}= 69.1\,\rm{km/s}$, and $w_{\perp, \mathrm{core}}= 121.2\,\rm{km/s}$. The beam is defined at a parallel drift velocity of $89.8\,\rm{km/s}$ from the core with $n_{\mathrm{beam}} = 236.1\,\rm{cm}^{-3}$, $w_{\parallel, \mathrm{beam}}= 149.9\,\rm{km/s}$, and $w_{\perp, \mathrm{beam}}= 142.2\,\rm{km/s}$. With this form, we can test how well g-SBR reconstructions recover both the core and beam structure and the statistical moments of the whole VDF. 

% With the model VDF defined, we define a magnetic field and bulk plasma velocity vector in the instrument frame, then transform the distribution from the plasma frame into the instrument frame. We interpolate the resulting distribution linearly onto the discrete grids, using the defined energy, azimuth, and elevation angles. This interpolated distribution serves as our synthetic measurement. 
We evaluate our model VDF on the synthetic ESA instrument grids, which are based on the SPAN-Ai grids \citep{Livi_etal_2022}. The synthetic instrument grids have 32 energy channels, 8 elevation, and 8 azimuth angles such that $\theta_{\mathrm{grid}} \in [-52^{\circ}, 52^{\circ}]$, $\phi_{\mathrm{grid}} \in [174^{\circ}, 95^{\circ}]$, and $E_{\mathrm{grid}} \in [21~\textrm{eV}, 17616~\textrm{eV}]$. These points are the grid centers on which the model is evaluated. Next, we define the magnetic field direction $\boldsymbol{\hat{b}}_{\mathrm{model}}$ (in the instrument frame) and the plasma bulk velocity $\vec{u}_{\mathrm{model}}$ to transform the instrument grids into the field aligned coordinates that we plug into Eqn.~(\ref{eq:biMax}). 

To validate the reconstruction methods under different observing geometries, we define two tests where we systematically vary the alignment of the magnetic field and bulk plasma velocity relative to the instrument frame. For each configuration, we compute the velocity moments from the reconstructed distribution and compare them to the moments of the full model VDF.

\begin{figure*}
    \centering
    \includegraphics[width=\linewidth]{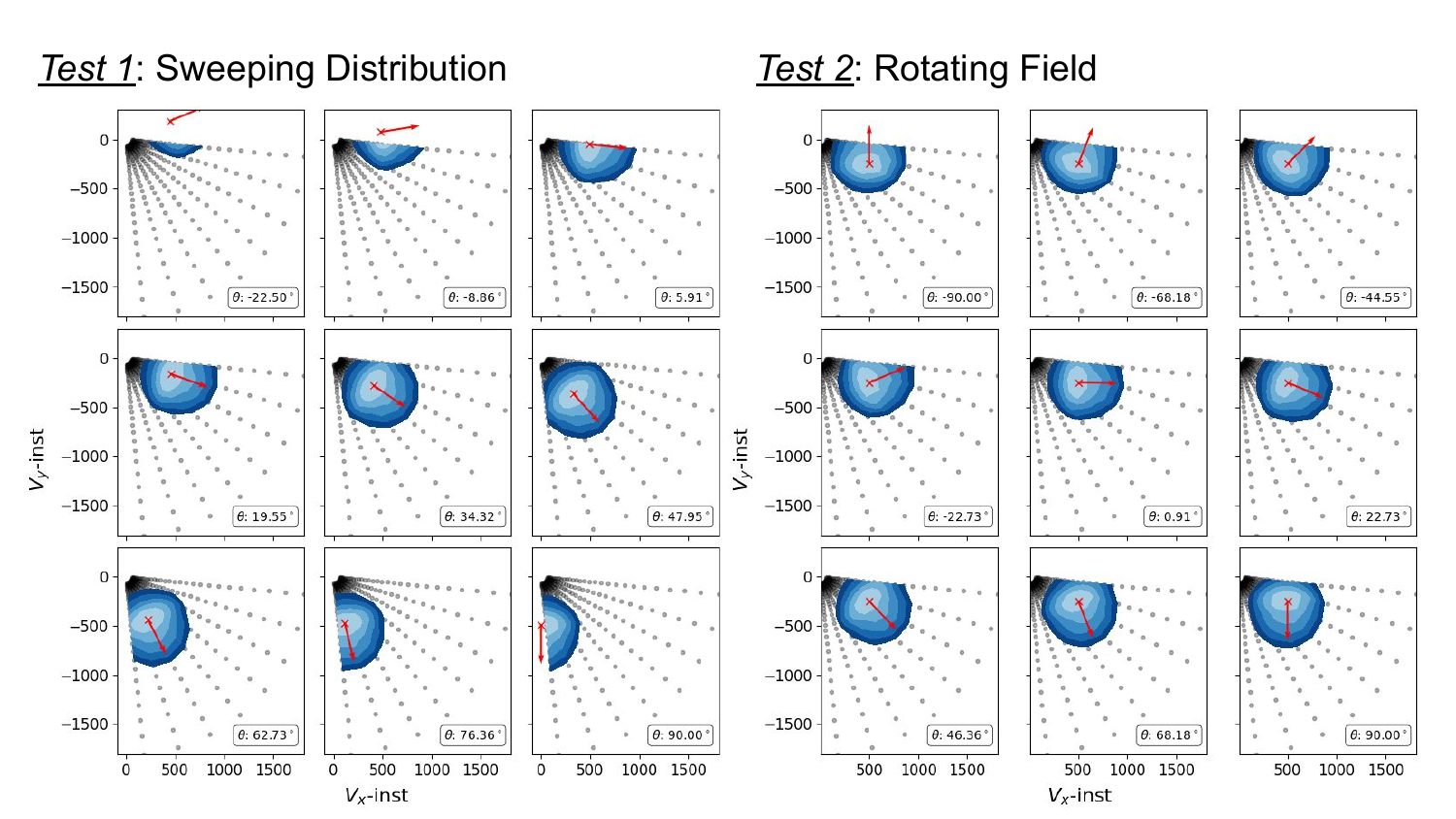}
    \caption{Demonstrations of the two test cases used in this study. Blue contours of the model distribution function are plotted on the $x$-$y$ instrument plane, with the direction of the magnetic field $\boldsymbol{\hat{b}}_{\mathrm{model}}$ denoted by the red arrow, centered at the model bulk speed $\mathbf{u}_{\mathrm{model}}$. The black dots represent the grid centers of the synthetic ESA instrument grid, inspired by SPAN-Ai. The \textit{left} nine panels show select slices for the sweeping distribution test as a function of magnetic field angle $\theta_{\mathrm{rot}}$. Here, $\theta_{\mathrm{rot}}$ goes from $-22.5^{\circ}$, behind the heat-shield to $+90^\circ$ degrees, end of prime FOV. The \textit{right} nine panels show select slices for the rotating field test in which the model bulk speed is set to be $\mathbf{u}_{\mathrm{model}} = (500, -250, 0)$, and $\theta_{\mathrm{rot}}$ goes from $-90^\circ$, behind the heat shield, to $+90^\circ$. Note that in the rotating field test, we completely lose the beam for all angles less than $-50^\circ$. }
    \label{fig:bimax_tests}
\end{figure*}

\textsc{Test 1}: \textit{Sweeping Distribution.} We perform a simultaneous azimuthal rotation of the magnetic field and bulk velocity vector. 
%, sweeping from just behind the heat shield toward the \(-\hat{y}\) direction in the SPAN-Ai instrument frame. 
The magnetic field unit vector $\boldsymbol{\hat{b}}_{\mathrm{model}}$ rotates in the $x$-$y$ instrument plane by an angle defined from the $+\hat{x}$-axis, such that $\theta_{\mathrm{rot}} \in (-22^{o}, 90^{o})$. The model bulk speed is defined to be $\mathbf{u}_{\mathrm{model}}\,[\mathrm{km/s}] = 500 \, \boldsymbol{\hat{b}}_{\mathrm{model}}$, so that the magnetic field and bulk speed rotate in lockstep relative to the instrument. 
The left panel of Fig.~\ref{fig:bimax_tests} shows nine example slices of the model VDF rotation. At $\theta_{\mathrm{rot}} \sim 0^\circ$, just below $50\%$ of the total distribution is measured. This test examines how far a VDF can lie outside the instrument's FOV while still allowing the g-SBR method to produce an accurate gyrotropic reconstruction. 

\textsc{Test 2}: \textit{Rotating Field.} We fix the core velocity and carry out an azimuthal rotation of the magnetic field vector. For this case, the core velocity is $\mathbf{u}_{\mathrm{model}}\,[\mathrm{km/s}] = 500 \,\boldsymbol{\hat{x}} + 250\,\boldsymbol{\hat{y}} + 0 \,\boldsymbol{\hat{z}}$. The magnetic field again rotates in the $x$-$y$ plane, such that $\theta_{\mathrm{rot}}$ rotates from $-90^{o}$ to $90^{o}$ with respect to the $+\hat{x}$-direction. When $\theta_{\mathrm{rot}}$ is negative, the beam of the bi-Maxwellian model is partially or completely obstructed by the heat shield, with only the core of the distribution being measured. The test is shown in the right panels of Fig.~\ref{fig:bimax_tests}.

\subsection{Instrument Grid Moments}\label{sec:grid_moms}

To calculate the moments as they would be calculated in an ESA data, we evaluate the model distribution on the spherical instrument grids in velocity space, defined by the energy E (or equivalently energy–derived speed $v = \sqrt{2\,q\,E / m}$), the polar angle $\theta$, and the azimuth $\phi$ \citep{Paschmann_Moments}. The instrument's energy, polar angle, and azimuthal angle define the instrument grid centers, from which we define the bin edges, with logarithmic spacing in $v$ and linear spacing in $\theta$ and $\phi$. The velocity–space volume element is given by 
\begin{equation}
    \Delta V^{ijk} \;=\; (v^i)^2 \, \sin\theta^j \, \Delta v^i \, \Delta\theta^j \, \Delta\phi^k \, .
\end{equation}
The velocity components at the bin centers are
\begin{eqnarray}
    v_{x}^{ijk} &=& v^i \sin\theta^j \cos\phi^k \, , \\
    v_{y}^{ijk} &=& v^i \sin\theta^j \sin\phi^k \, , \\
    v_{z}^{ijk} &=& v^i \cos\theta^j \, .
\end{eqnarray}

The moments are evaluated as discrete Riemann sums over the grid, i.e. 
\begin{eqnarray}
    n &=& \sum_{ijk} f^{ijk} \, \Delta V^{ijk} \, , \\
    u_{\alpha} &=& \frac{1}{n} \sum_{ijk} v_{\alpha}^{ijk}\, f^{ijk} \, \Delta V^{ijk} 
    \quad \text{for } \alpha \in \{x,y,z\} \, , \\
    T_{\alpha\beta} &=& \frac{m}{k_B n} \sum_{ijk} 
    \left( v_{\alpha}^{ijk} - u_{\alpha} \right)
    \left( v_{\beta}^{ijk} - u_{\beta} \right) 
    f^{ijk}\, \Delta V^{ijk} \, , \nonumber\\
\end{eqnarray}
where $f_{ijk}$ is the VDF value evaluated at the grid center, $n$ is the number density, $\mathbf{u} = (u_x,u_y,u_z)$ is the bulk velocity, and 
$\mathsf{T}$ is the symmetric temperature tensor. The scalar temperature is obtained from the trace of $\mathsf{T}$,
\begin{equation}
    T = \frac{1}{3} \, \mathrm{tr}\,\mathsf{T} \, .
\end{equation}
Unless otherwise specified, conventional solar wind units are assumed with $n$ in $\mathrm{cm^{-3}}$, $\mathbf{u}$ in $\mathrm{km\,s^{-1}}$, 
and $T$ in $\mathrm{K}$.

\begin{figure*}
    \centering
    \includegraphics[width=0.49\linewidth]{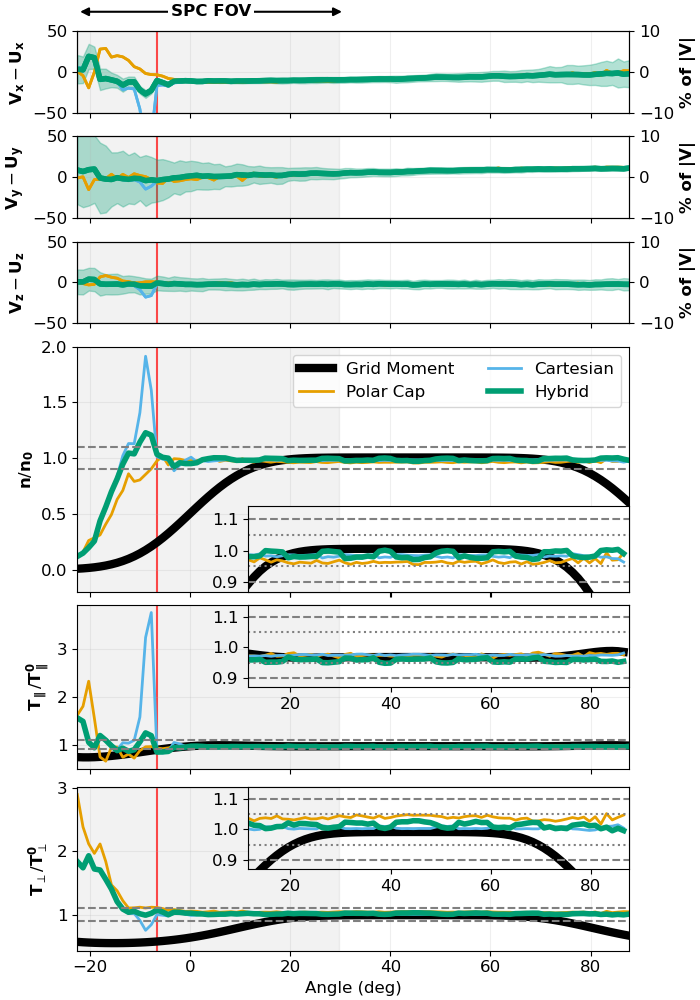}
    \includegraphics[width=0.49\linewidth]{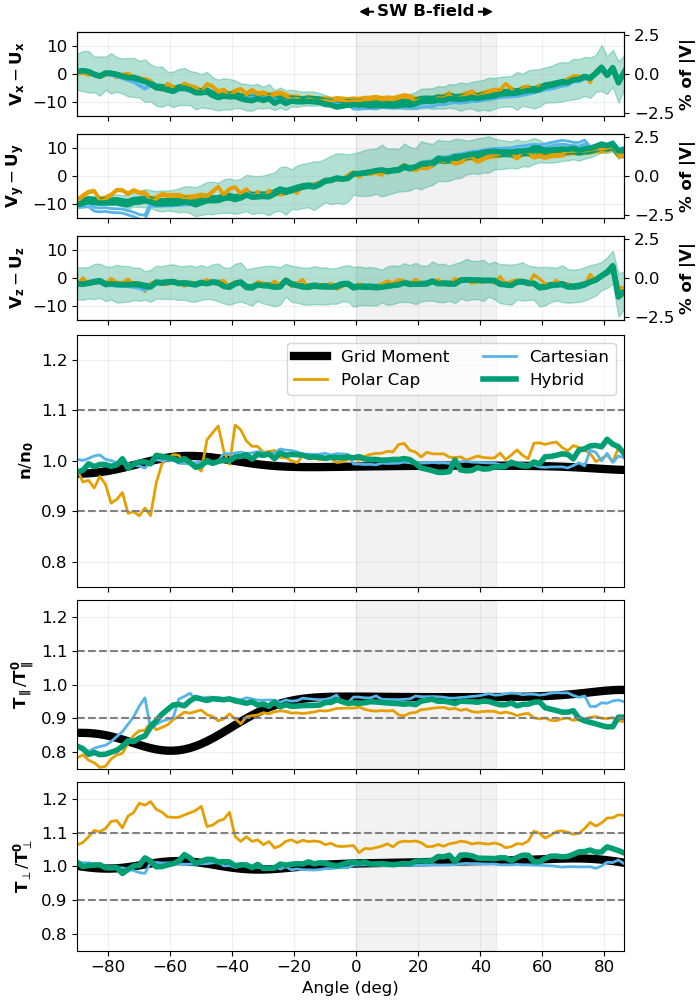}
    \caption{Comparison of velocity, density, and temperature moments for \textsc{Test~1} (\textit{left} panels) and \textsc{Test~2} (\textit{right} panels). In \textsc{Test~1}, the magnetic field vector and bulk flow direction co-rotate in angle around the instrument frame, while in \textsc{Test~2}, the velocity vector is fixed and the magnetic field rotates about the instrument frame origin. In both cases, the partial moments computed from the synthetic ESA grids are shown in \textit{black}. Gyrotropic moments computed using three different 2D reconstruction methods --- polar cap (orange), Cartesian (cyan), and hybrid (green)---are shown for comparison. The top three panels show the reconstructed velocity components differences with respect to the corresponding model bulk velocity components $V_x - U_x$, $V_y - U_y$, and $V_z - U_z$. The fourth panel presents the total number density in $\mathrm{cm}^{-3}$, with the dashed purple line indicating the model density of $1703.0~\mathrm{cm}^{-3}$. The bottom two panels show the parallel and perpendicular temperature components, normalized to the model temperatures $T_{\parallel}^0 = 573{,}551.14~\mathrm{K}$ and $T_{\perp}^0 = 943{,}147.10~\mathrm{K}$, respectively. In the left panels, the gray shaded regions indicate the approximate field of view (FOV) of the SPC instrument. Insets in selected panels highlight features in the central angular region. The gray-shaded region for the right panels represents the typical ranges of solar wind conditions in the inner heliosphere.  
}
    \label{fig:moment_comparison}
\end{figure*}

\subsection{Moment Comparison} \label{sec:moment_comparison}

Figure~\ref{fig:moment_comparison} shows the reconstructed plasma moments for both the sweeping distribution test (\textit{left} panels) and the rotating field test (\textit{right} panels), using the three SBR approaches: the polar cap method (\textit{orange}), the Cartesian method (\textit{cyan}), and the hybrid method (\textit{green}). We evaluate the moments for each of these methods using the super-resolved GDF on high-resolution grids. For all cases shown here, we use a $49 \times 49$ grid in $(v_{\perp}, v_{\parallel})$ phase-space, which is selected to adequately resolve the VDF and support convergence of the moment integrals calculations laid out in Section~\ref {sec:gdf_moms}. For clarity, we define the terminology used in this section as follows: the known, model-derived plasma moments are referred to as \textit{model moments}; the Riemann-integrated moments, representative of those obtained from electrostatic analyzers (ESAs), as \textit{instrument moments}; and the moments derived from the g-SBR method as \textit{reconstructed moments} or \textit{g-SBR moments}.

For each velocity moment, we evaluate how well the reconstruction preserves the corresponding bi-Maxwellian model moments. The first three panels show the velocity components, with both the signed difference from the model bulk velocity ($v_{i,\mathrm{rec}} - u_{i,\mathrm{model}}$) and the signed percent difference in the velocity magnitude, normalized by the model bulk speed, i.e. $(v_{i,\mathrm{rec}} - u_{i,\mathrm{model}})/|\mathbf{u}_{\mathrm{model}}|$. The fourth panel compares the reconstructed density $n_{\mathrm{rec}}$ with the model density $n_{0} = 1702~\mathrm{cm}^{-3}$. Finally, the fifth and sixth panels show the reconstructed parallel and perpendicular temperatures, each normalized by their respective model values, $T_{\parallel}^0 = 573{,}551.14~\mathrm{K}$ and $T_{\perp}^0 = 943{,}147.10~\mathrm{K}$. In panels four through six, we plot the instrument moments (\textit{black} curves), derived by integrating the model distribution over the simulated measurement grid domain, using the Riemann sum described in Section~\ref{sec:grid_moms}.

% To accurately transform into the plasma frame, we employ the gyro-centroid optimization routine (see Sec.~\ref{sec:gyro-centroid-finder}), which refines the initially reported bulk velocity by minimizing the data misfit by accounting for axisymmetry about $\boldsymbol{\hat{b}}_{\mathrm{model}}$. This procedure both identifies the correct symmetry axis (gyroaxis) and provides a consistent estimate of the bulk velocity uncertainty. Because these uncertainties arise solely from the centroid-finding process, they are identical across all three reconstruction methods (polar cap, Cartesian, and hybrid). For visual clarity, we present only hybrid reconstruction padded with the associated uncertainties (shown as shaded green contours), while noting that these uncertainty bounds apply equally to the other methods.
We use the gyro-centroid optimization routine (see Sec.~\ref{sec:gyro-centroid-finder}) to transform the distribution into the plasma frame. This process identifies the gyroaxis and provides estimates of bulk velocity uncertainty, which are the same across the three reconstruction methods: polar cap, Cartesian, and hybrid. For clarity, we show only the hybrid reconstruction with uncertainties represented as shaded green contours, noting that these bounds also apply to the other methods.

\begin{figure*}
    \centering
    \includegraphics[width=\linewidth]{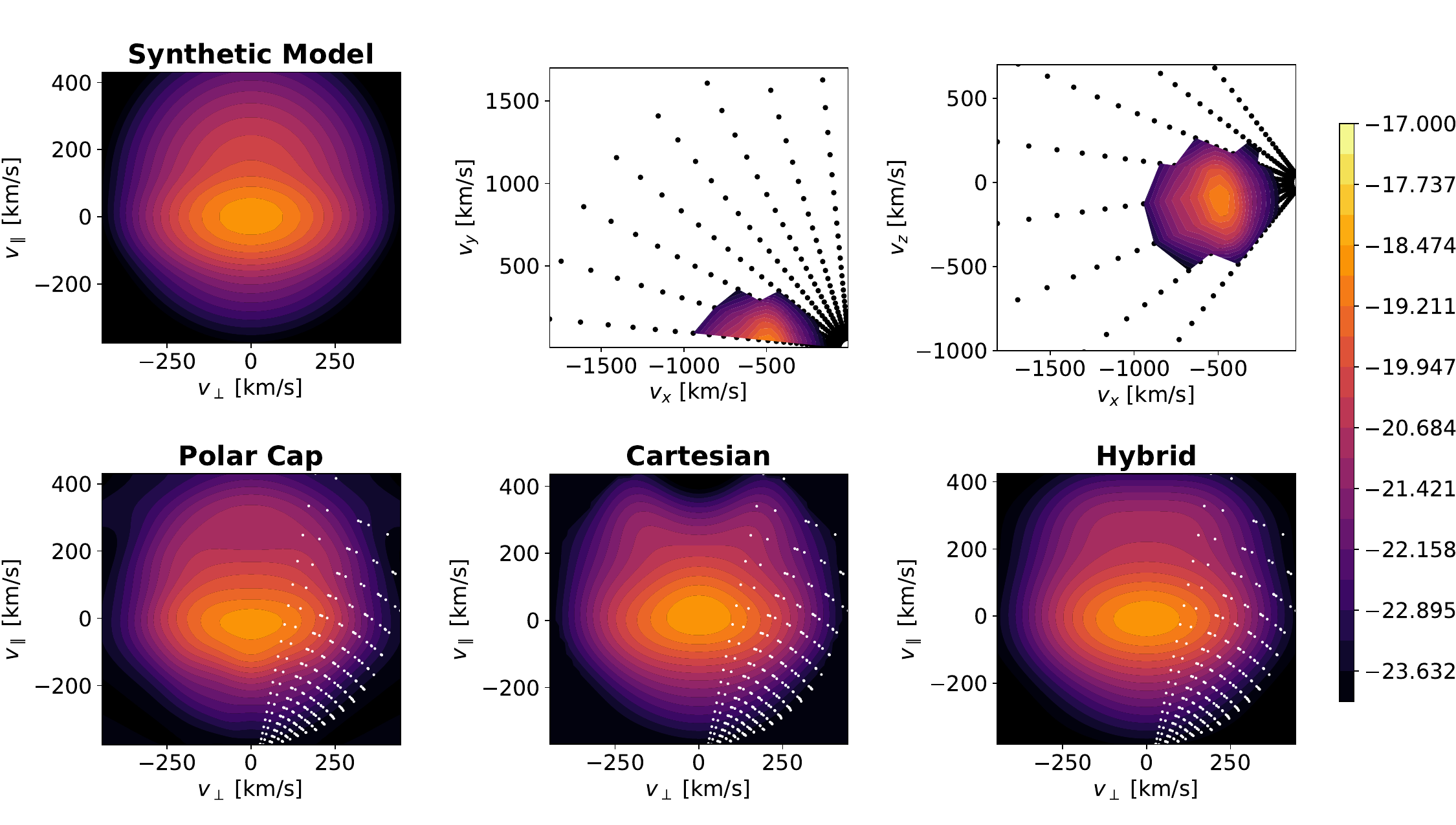}
    \caption{The \textit{top} panel shows the synthetic model and the corresponding $v_x-v_y$ and $v_x-v_z$ instrument grid slices for the case demarcated by the \textit{red} vertical line of \textsc{Test 1} shown in Fig.~\ref{fig:moment_comparison}. This is the case where the beam edge is captured at a grazing angle by the ESA grids. The measured partial density moments accounts for $\sim 25\%$ of the total density, indicating that both the beam and the (dominant fraction of) core is outside the FOV. This is the first point where all the g-SBR methods polar cap, Cartesian, and hybrid, recover the synthetic model moments. The \textit{bottom} panels show the final reconstruction for each of the methods. Despite completely missing the beam, the polar-cap Slepians recover the beam but have small, low-amplitude wiggles at low $v$. The Cartesian Slepians provide a clean reconstruction at low $v$ but fail to recover the beam. The hybrid method recover both the beam (due to support provided by the polar cap method) and have a clean low energy reconstruction (due to support provided by the Cartesian method). The \textit{white} dots in the lower panels indicate the location of all the FAC grids which have a significant count to be considered for fitting.}
    \label{fig:grazing-angle-demo}
\end{figure*}

Inspecting the sweeping case (panels in the \textit{left} column), we can subdivide the sweeping angle $\theta$ = $\cos^{-1}{\left(\boldsymbol{\hat{b}}_{\mathrm{model}} \cdot \boldsymbol{\hat{x}}\right)}$ axis into three broad regimes. The first regime includes magnetic field angles ranging from $-22.5^{\circ}$ to approximately $0^{\circ}$. Examples of the measured distribution are reported in the first row of Fig.~\ref{fig:bimax_tests}. In this region, the bulk speed $\mathbf{u}_{0}$ falls outside the instrument FOV. This region serves as a diagnostic for identifying the minimum amount of information needed to reconstruct the ion distribution function. Looking at the first three panels of Fig.~\ref{fig:moment_comparison} in this range, there are differences from the model velocities, on the order of $30~\mathrm{km/s}$ for the $v_{x}$ component alone, and the $v_y$ and $v_z$ components are around $10~\mathrm{km/s}$. However, as the magnetic field rotates toward $0^{\circ}$, all the velocity components converge to within $5~\mathrm{km/s}$ of the actual model velocity. This point at which the g-SBR methods converge to the model velocities is demarcated by the red vertical line in Fig.~\ref{fig:moment_comparison}. The convergence depends on whether or not there are sufficient number of grids constraining the model parameters (Slepian basis). We see that the moments start converging around the point where the magnetic field and the nearest field-aligned grids are within the angular resolution $L_{\mathrm{max}}$ of the Slepian basis. This is demonstrated later in Fig.~\ref{fig:grazing-angle-demo}.
Looking at the uncertainty in the three velocity components, the variations in $v_x$ and $v_z$ remain well constrained, while $v_y$ has variations up to $50~\mathrm{km/s}$. This is because the distribution is FOV-limited in the $+\boldsymbol{\hat{y}}_{\mathrm{inst}}$ direction. The $v_x$ and $v_z$ components remain largely unaffected by the obstruction, as the distribution is mainly unaffected in the $+\boldsymbol{\hat{x}}$ and $+\boldsymbol{\hat{z}}$ directions. However, limited constraints in the $+\boldsymbol{\hat{y}}$ direction allow more freedom, thus higher variance, in the Bayesian centroid approximation. It is important to note that the reconstructed velocity moments do not vary significantly between the three methods beyond the vertical red line. This is expected as the velocity moment is strongly weighted towards the peak of the distribution. As long as the three g-SBR methods converge to stable solutions, the velocity is accurately recovered.  

The density is shown in the fourth panel in the \textit{left} column of Fig.~\ref{fig:moment_comparison}. In the region with magnetic field angles from $-22.5^{\circ}$ to $0^{\circ}$, we see that despite resolving the velocity moment within $10\%$ of the velocity magnitude, the density differs significantly from the model density. The instrument density moment illustrates the impact of the partial FOV: as the distribution rotates into the prime FOV, the recovered density increases. In comparison, the polar cap, Cartesian, and hybrid g-SBR methods all converge to the underlying model density much more rapidly, indicating that these methods are accounting for the missing structure that is FOV obstructed. However, recovering the plasma density alone does not ensure that we have recovered the higher-order moments of the distribution. 
% At the vertical \textit{red} line, the point at which the velocities converged, the conventionally measured density is $24.4~\%$ of the actual density $n_{0} = 1702~\textrm{cm}^{-3}$, is at an angle of $-6.6^\circ$ which roughly corresponds to a measured distribution plotted in the upper middle panel of the Sweeping Distribution panel in Fig.~\ref{fig:bimax_tests}. 
At the vertical \textit{red} line, which marks the point where the velocities converge, the instrument density is $24.4~\%$ of the model density, $n_{0} = 1702~\textrm{cm}^{-3}$. The line is defined at an angle of $\theta = -6.6^\circ$ and corresponds to a measured distribution that resembles the one shown in the upper middle panel of Fig. \ref{fig:bimax_tests}.
At this point, all three g-SBR methods come within $\geq 95\%$ of the model density. At $-15^{\circ}$, the Polar Cap method estimates a density that is $85\%$ of the total model density. This, in fact, corresponds to the core model density of $n_{\mathrm{core}} = 1466.9~\mathrm{cm}^{-3}$. In this configuration, the FAC grids provide no constraint along the field-parallel direction, providing poor support for the beam. There is a spike in the Cartesian density here, where the beam is poorly constrained. This is due to the grid separation at high $v$ is larger than the maximum supported wavenumber $k_{\mathrm{max}}$ (used to derive the maximum $N^{\mathrm{2D}}_{\mathrm{cart}}$ as in Eqn.~[\ref{eqn:N2D_cartesian}]). The Cartesian method is therefore poorly conditioned to this regime, since the regions between gridpoints near the beam fall in the null space of the inverse problem. If the ``beam'' were removed from the moment calculation, the Cartesian method would reduce to the model's core density. The hybrid method, however, does not suffer from this limitation, as the polar cap method provides a model constraint that stabilizes the beam. While the hybrid reconstructed density at the vertical \textit{red} line is $10\%$ larger than the model density, the hybrid reconstruction remains well-conditioned. 

In panels five and six, similar trends are observed in the temperature reconstructions for the same angular region.  The instrument parallel temperature $T_{\parallel}$ (\textit{black} curve in panel five) is a systematic underestimate at low rotation angles because the beam is not captured on the instrument grids. During that interval, the instrument parallel temperature is approximately $70\%$ of the model distribution's actual moment. Furthermore, the instrument perpendicular temperature $T_{\perp}$ (black curve in panel six) is only $60\%$ of the model's reported perpendicular temperature moment. This is also due to the FOV obstruction; even the g-SBR reconstructions, before the vertical \textit{red} line, show significant variations in that moment estimate. However, for this model VDF, as soon as $24.4\%$ of the distribution overlaps with the instrument grids, all three g-SBR methods converge to $\geq 95\%$ of the model temperatures.  

Overall, this suggests that velocity is well-approximated as long as \textit{some} measurements are taken on the instrument grids (particularly $v_x$ and $v_z$). For this type of core-beam model VDF, when the instrument grids captured less than $24.4~\%$ of the VDF by density, the g-SBR method only approximated the core of the distribution. When the instrument FOV is sufficient to constrain the linear inversions, however, the velocity components are estimated to within $5~\mathrm{km/s}$ of actual, and both the density and two temperature components are within $\geq 95~\%$. 

For $0^{\circ}$ to $65^{\circ}$, it is clear from the second panel that all three g-SBR methods converge to a stable solution. This is to be expected because at least half of the distribution is within the instrument FOV. In this angle range, the conventional instrument moments also converge to the model density and both model temperature components. 
% However, it is essential to note that the density and perpendicular temperature \textit{typically} are underestimated by the conventional moments unless the peak of the distribution is \textit{entirely} inside the FOV 
However, it is essential to note that the density and perpendicular temperature from the instrument moments are not fully constrained unless the peak of the distribution lies \textit{entirely} within the FOV (i.e., the lightest blue contour in the left middle panel of Fig.~\ref{fig:bimax_tests}), which often results in underestimation, though the exact bias depends on the sampled region of phase space. In the region from $65^{\circ}$ to $90^{\circ}$, the instrument moments drop as the distribution begins to rotate outside the FOV in the $-\boldsymbol{\hat{x}}$ direction. However, the g-SBR methods remain true to the model parameters, as more than $24.4\%$ of the total density is measured on the instrument grids. The zoomed-in insets in panels four through six show a small but clear periodic pattern that emerges in all three methods. This periodicity comes from the fine alignment of the magnetic field with the instrument grids. The tiny dips in this periodic structure occur when the magnetic field rotates between two sets of radial grid branches.

To quantify the impact of the instrument orientation relative to the B-field, we turn to \textsc{Test 2}. In this test, the bulk velocity is anchored to $\mathbf{u} = (500, -250, 0)~[\mathrm{km/s]}$. The magnetic field rotates from the $+\boldsymbol{\hat{y}}$ to $-\boldsymbol{\hat{y}}$, as shown in Fig.~\ref{fig:bimax_tests}. In this test, the peak of the distribution is always within the instrument FOV, but the beam falls outside the FOV to varying degrees. We therefore expect the partial moment density and the g-SBR densities to match the model density closely in all cases. The beam accounts for $\sim 14~\%$ of the total model density; therefore, we expect around a $\sim 10\%$ variation in the reconstruction and instrument densities when the beam is cut off. 

For \textsc{Test 2}, the beam component is partially or totally occulted for sweep angles from $-90^{\circ}$ to $-45^{\circ}$ (top panel of \textit{right} half of Fig.~\ref{fig:bimax_tests}). The instrument-reported density (\textit{black} curve) is systematically lower than the model-reported density in this range, but it falls within $95~\%$ of the actual value. There is a noticeable increase in instrument density at $-60^\circ$, which is attributed to the interpolation scheme that is agnostic to the model, resulting in slight overestimates when integrated. The Cartesian and hybrid g-SBR moments agree with the model moments. The partial field of view effect is pronounced, however, in the parallel temperature estimates from panel five. Though the instrument moments recover the proper density, there is a systematic underestimate of the parallel temperature throughout the angle range from $-90^{\circ}$ to $\approx -35^{\circ}$. The full parallel core + beam structure is not resolved on the instrument grids. We see that within the first $30^{\circ}$ of this sub-region, both the hybrid and Cartesian methods also underestimate the parallel temperature. Once $-60\%$ is reached, the models converge to $\ge 95\%$ of the actual value. Finally, the perpendicular temperature is well estimated for all but the polar cap g-SBR method.

Once the beam is resolved in the instrument FOV, the instrument moments and the reconstructed moments all produce estimates within $90\%$ of models. There is a systematic variation in the velocity estimators, particularly in the $v_x$ and $v_y$ components, which is cyclic with respect to the alignment angle.
% This is solely related to the discrete angular and energy resolution of the instrument grids, as the peak of the distribution is anchored between several grids. 
This is related to the peak of the instrument measured distribution falling between multiple discrete angular and energy grids.
The same trends observed in the three reconstructions are also evident in the instrument velocities. When compared to the magnitude of the velocity $|v| \approx 560~\mathrm{km/s}$, the variation in the velocity components is on the order of $2.5\%$. The variation in the velocity moment estimate is also dependent on the observed energy. The higher the solar wind speed, the larger the uncertainty and variation in the plasma moments.

The Polar Cap method is an extension of the agyrotropic approach presented in \cite{Das_Terres_2025}, and that method is particularly well-suited for cases where the magnetic field and bulk velocity are nearly aligned (as seen from test 1: the sweeping case). 
% In this regime, the gyro-frame grids become almost aligned with respect to the magnetic field, 
In this regime, the gyro-frame grids are nearly identical to the intrinsic instrument grids, 
allowing polar caps to be constructed on each energy shell with constant angular spacing. It is therefore unsurprising that the polar cap method performs well for the sweeping test: the gyroframe grids trace out radial contours, and the data are symmetrically represented. That said, while the polar cap method can also reconstruct the distributions in \textsc{Test 2}, the resulting plasma moments show larger variations compared to the Cartesian and hybrid methods. 
% When the field-aligned instrument grids are irregularly sampled, as they are in \textsc{Test 2}, it is more difficult to define the log-spaced knots. 
When the field-aligned instrument grids do not fall on distinct radial shells, and are irregularly sampled, as they are in \textsc{Test 2}, it is more difficult to define the location of the log-spaced knots.
As a result, although the fits recover density, temperature, and velocity within approximately $10\%$, \textsc{Test 2} highlights the limitations of the polar cap method when the magnetic field is not aligned radially with the instrument grids configuration.

The hybrid method is robust to challenges presented by both \textsc{Test 1} and \textsc{Test 2}. As shown above, the Cartesian method can be problematic in cases where the beam is poorly supported (\textsc{Test 1}), resulting in spikes in density and temperature due to the lack of field-aligned constraints. Conversely, the polar cap method struggles with irregularly sampled grids (\textsc{Test 2}), leading to larger variations in plasma moment estimates. When either method is applied under favorable conditions, it recovers the underlying distribution with good accuracy: Cartesian excels when the beam curvature enters the FOV (e.g., $\sim -45^{\circ}$ in \textsc{Test 2}) and polar cap, by construction, enforces radial symmetry when the field and flow are aligned. The strength of the hybrid approach is that it combines these advantages—utilizing the polar cap approach to robustly constrain the beam while retaining the Cartesian method flexibility to handle irregular sampling at low $v$. As a result, the hybrid method reproduces the model distribution very well across both tests, capturing the plasma moments and kinetic structure with high fidelity (see Fig.~\ref{fig:grazing-angle-demo}).

In general, the qualitative nature of Fig.~\ref{fig:moment_comparison} holds for more anisotropic or higher density beams. However, the exact angular extent at which the moments converge varies on a case-by-case basis, i.e., the direction of the magnetic field, the thermal extent of the ion populations, and the gyro-centroid. In the case of \textsc{Test 1}, we would expect the convergence to always fall within the $L_\mathrm{max}$ condition described earlier.
 
\section{Discussions and future work}

This is the second paper in the Slepian Basis Reconstruction (SBR) series, in which we extend the ability to super-resolve discrete plasma measurements to a robust gyrotropic estimate. In \textsc{Paper I} \cite{Das_Terres_2025}, we introduced the Slepian functions and the SBR method in the context of parameterizing VDFs. Our focus in \textsc{Paper I} was on ESA measurements collected onboard the Magnetospheric Multiscale (MMS) mission and Solar Orbiter, where VDFs are usually not subject to FOV limitations. Full coverage enabled fully agyrotropic reconstructions, which are particularly relevant for the hot magnetospheric plasma measured by MMS. Solar wind VDFs have a much lower Mach number and are generally measured over an accumulation time much longer than the ion gyro-period, which justifies the gyrotropic assumption.  
%We present our second paper in a series of papers where we use the Slepian Basis Reconstruction (SBR) method to characterize and super-resolve solar wind VDFs from discrete measurements. In \textsc{Paper I} \cite{Das_Terres_2025}, we introduced the Slepian functions and the SBR method in the context of parameterizing VDFs. In \textsc{Paper I}, we had focused on ESA measurements onboard MMS and Solar Orbiter where VDFs are not subject to FOV limitations. Having access to the complete target region in the solid angle space (azimuth-elevation) had further enabled us to perform fully agyrotropic reconstructions. This is especially relevant for high cadence MMS measurements in the hot magnetosphere where gyrotropy is often not an accurate approximation. For solar wind VDFs, however, when comparing the accumulation time of the ESA against the ion gyro-periods, invoking gyrotropy is a dominantly safe assumption.

In this study (\textsc{Paper II}), we have extended the SBR method to produce robust gyrotropic VDF reconstructions that carefully account for the instrument grid resolution in the field-aligned gyrotropic frame. We devised the mathematical formalism g-SBR for inferring a continuum representation of the gyrotropic distribution from discrete ESA measurements. The g-SBR approach effectively constrains VDFs that meet the following criteria: (A) they are dominantly gyrotropic, (B) they are measured on a relatively large accumulation time compared to the gyro-period, and (C) they are limited by instrumental restrictions. For the last case, the g-SBR method allows us to recover the occulted portion of the VDF via symmetry considerations. In keeping with the best practices of scientific computation and open-source reproducible code development, we provide our Python-based\footnote{Note that the current version has a wrapper around \textsc{Matlab} source codes for generating Slepian functions which uses \cite{slepian_alpha} and \cite{slepian_foxtrot}} \texttt{gdf} module along with extensive documentation hosted on GitHub pages.
% In this study (\textsc{Paper II}), we extend the SBR method to produce a robust gyrotropic reconstruction while carefully accounting for the instrument grid resolution in the field-aligned gyrotropic frame. This mathematical formalism g-SBR of inferring a continuum representation of the gyrotropic distribution from discrete ESA measurements developed here is devised to be applicable to (A) where the underlying VDF is expected to be dominantly gyrotropic, (B) where the solar wind conditions are conducive to accommodating multiple ion gyro-periods within the ESA accumulation time $\tau_{\mathrm{gyro}} \gg \tau_{\mathrm{acc}}$, and (C) instruments where the ESA has FOV restrictions, thereby allowing us to recover the occulted portion of the VDF via symmetry considerations. In keeping with the best practices of scientific computation and open-source reproducible code development, we provide our Python-based\footnote{Note that the current version has a wrapper around \textsc{Matlab} source codes for generating Slepian functions which uses \cite{slepian_alpha} and \cite{slepian_foxtrot}} \texttt{gdf} module along with elaborate documentation hosted on GitHub pages.

The novelty of this paper lies in the three mathematical model frameworks developed to represent a gyrotropic distribution: polar cap, Cartesian, and hybrid. Through the synthetic tests, we highlight the practical advantages of using each method. In real-life applications, the grids projected in the gyrotropic FAC change at each time-stamp, depending on the magnetic field orientation and the bulk velocity of the VDF. This gives rise to several questions when formulating a reconstruction/inverse problem: 
\begin{itemize}[noitemsep,topsep=0pt,wide,labelindent=5pt]
    \item Do the FAC grids retain their inherent spherical symmetry?
    \item If not, can the positions of the arbitrary unstructured grids be connected to the instrument's angular resolution to yield a well-conditioned inverse problem? 
    \item If not, are the grid measurements at the low energy end of the spectra too irregular and noisy, resulting in a highly granular representation at low energies? 
    \item Is the dominant part of the distribution too FOV-restricted to allow a meaningful reconstruction unless complemented by independent instruments such as a Faraday Cup (for instance, adding in the Solar Probe Cup measurements to SPAN-Ai for PSP data)?
\end{itemize}
% The primary novelty of this paper are the three mathematical model frameworks developed to represent a gyrotropic distribution --- polar cap, Cartesian and hybrid. Section~\ref{sec:methods} goes over each of these methods. Through the synthetic tests in Section~\ref{sec:synthetic_demo}, we highlight the respective practical advantages of using each method. The grids, projected in the gyrotropic FAC changes at each time-stamp depending on the magnetic field orientation as well as the bulk velocity of the VDF. This presents the following questions when formulating a reconstruction/inverse problem (i) Do the FAC grids retain its inherent spherical symmetry?, (ii) If not, can the positions of the arbitrary unstructured grids be connected to the instrument's angular resolution to yield a well-condition inverse problem? (iii) If not, are the grid measurements at the low energy end of the spectra too irregular and noisy resulting in highly granular representation at low energies, (iv) is the dominant part of the distribution too FOV restricted to allow a meaningful reconstruction unless complemented by independent instruments such as a Faraday Cup (for instance, adding in the Solar Probe Cup measurements to SPAN-Ai for PSP data).

Each method in Section~\ref{sec:methods} has been developed with careful attention to tackle different aspects of these problems. The polar cap method (described in Sec.~\ref {sec:polcap}), which uses polar-cap based Slepian functions, is particularly suitable for reconstructions when the magnetic field is dominantly oriented along the instrument's radial direction. Polar cap Slepians are defined in angular space and therefore not susceptible to large $v_{\perp}$ gaps at high energies, so long as the gap falls above $\Delta \phi$ the instrument resolution (or below the specified $L_{\mathrm{max}} = \pi / \Delta \phi$). This makes polar cap effective in reconstructing beams, even when the measurements only probe the beam edge. One of the downsides of polar cap is its large degree of freedom in the low-energy domain. This is because the knots in $v$ supporting the Slepian functions are log-spaced and therefore densely clustered at low $v$. The presence of spurious count bins that are poorly distributed may result in unphysical wiggles in the reconstruction at low $v$. 
% Based on private communications with our colleagues in the community, we have noted that this is a standard, well-recognized issue for all other reconstruction techniques. 
A common approach is to either apply an artificial smoothing (such as that imposed during fitting radial basis functions) or use a parametric fit, such as a bi-Maxwellian or Kappa distribution. 

We recognize that there is no reason to expect that the underlying ``true" VDF would be highly structured in some areas of the velocity phase space as compared to other regions (an automatic consequence of a simple instrument grid-based interpolation). Therefore, we introduce the Cartesian Slepian-based approach in Section~\ref{sec:cartesian}. Here, we estimate the minimum resolvable scales in phase space by accounting for FAC grid distributions. Consequently, the Cartesian method stipulates uniform granularity (maximum wavenumber $k_{\mathrm{max}}$) across all of phase space. It is easy to see that this naturally alleviates the drawback of wiggly low-energy reconstructions that persisted in the polar cap method. However, since Cartesian basis functions are defined in the velocity phase space and have no notion of the corresponding angular space, they do not share the advantage of being able to constrain the beam solely by measuring the beam edge. As a result, we see that the polar cap and the Cartesian approaches are highly complementary.

This led us to frame the hybrid method, which aims to combine the advantages of both these methods. Section~\ref{sec:hybrid} lays out the prescription where we essentially require the polar cap and Cartesian models to be jointly minimized, along with demanding their reconstructions to yield similar super-resolved GDFs. This results in an inverse problem where the polar cap basis renders stability to regions (such as the partially observed beam) that otherwise fall in the null space of the Cartesian basis functions, causing Cartesian to have high condition numbers. Conversely, the inversion also guarantees that the low-energy wiggles (which are below the maximum wavenumber resolution for Cartesian) never arise. Moreover, the structure of the polar cap and Cartesian bases are so complementary that we do not require an additional B-spline regularization (as was otherwise needed in polar cap, see Eqn.~[\ref{eqn:polcap_inv}]).  For visual reference, we add Fig.~\ref{fig:grazing-angle-demo} to show that hybrid is generally more robust than polar cap and Cartesian approaches using the case from \textsc{Test 1} where only the beam edge is captured at a grazing angle by the ESA measurements. The \textit{top} panel shows (from \textit{left} to \textit{right}) the synthetic underlying VDF model, the $v_x-v_y$ and $v_x-v_z$ slices of the ESA measurements that are used in our reconstruction. The \textit{bottom} panels show the corresponding three g-SBR reconstructions. We show that even when observed at a grazing angle, $polar cap$ can constrain the beam but is susceptible to low-amplitude wiggles at low $v$ knots. This is alleviated in the Cartesian reconstruction, which fails to construct the beam by nature of it falling in the null-space of the inversion. Finally, the hybrid method is shown to combine the advantages from both these methods --- it contains a beam while not being susceptible to unphysical wiggles at low $v$ grids. 

% Therefore, we recommend using hybrid as a default choice for all reconstructions since it was tested to yield robust results for arbitrary grid and magnetic field orientations.
Ultimately, we recommend using the hybrid method as a default choice for all reconstructions since it was tested to yield robust results for arbitrary grid and magnetic field orientations. Nonetheless, there are particular situations where polar cap or Cartesian methods may be better tailored than hybrid. The polar cap method provides a speed advantage in computations, making it ideal for quick or exploratory diagnostics. On the other hand, the Cartesian approach is better suited for preserving low \(v_{\parallel}\) structures within the ion distributions.

% Figure~\ref{fig:moment_comparison} in Sec.~\ref{sec:moment_comparison} uses two synthetic scenarios to benchmark the reconstruction efficiency using each of the three g-SBR methods. 
The two synthetic test scenarios in Sec.~\ref{sec:moment_comparison} serve as benchmarks for the reconstruction efficiency of the three g-SBR methods.
The choice of our model synthetic VDF and the magnetic field orientation are subsets of all possible configurations that could potentially arise in the solar wind measurements. However, the key takeaways from our synthetic tests can be summarized as
\begin{enumerate}[noitemsep,topsep=0pt,wide,labelindent=5pt]
    \item When at least half the VDF is in the FOV, all three methods in g-SBR perform adequately and preserve both macroscopic moments as well as kinetic structures.
    \item As the VDF goes outside the FOV, the partial moments from the instrument (such as density and temperature) are compromised systematically. However, depending on the FAC grid orientation, one or more of the g-SBR methods continue to preserve moments and kinetic structure. Of course, to what extent outside the FOV we can continue to expect the moment/structure preservation to hold depends on the underlying VDF's thermal extent.
    \item Unless drastically outside the FOV, g-SBR reports highly accurate $T_{\perp}$ and $T_{\parallel}$ estimates. This makes our method a demonstratively powerful choice to study temperature anisotropies in events such as switchbacks.
    \item Using the MCMC gyro-centroid refinement allows robust error estimates in velocity moments, which can be connected to FOV restrictions and the orientation of the plane perpendicular to $\boldsymbol{\hat{b}}_{\mathrm{model}}$. These are shown in the top three panels in Fig.~\ref{fig:moment_comparison}. In principle, future extensions of \texttt{gdf} may provide error estimates for the moment calculations by using the velocity uncertainties along with the magnetic field variance.
    \item The expected angle between the background magnetic field and the radial direction is in the range of $[0^{\circ}, 45^{\circ}]$ for typical solar wind conditions. The background field direction dictates how the instrument grids are transformed into the FAC frame. 
    % The expected range of solar wind conditions in the inner heliosphere boils down to the relative orientation of the magnetic field with the radial direction of the instrument coordinates being within $[0^{\circ}, 45^{\circ}]$. 
    We show that for cases where the VDF is squarely within the FOV, the irregularly spaced non-uniform grids are well handled by all the methods to yield accurate moments.
\end{enumerate}

The g-SBR formulation in this study, along with the \texttt{gdf} repository, adequately lays out the groundwork for adding in complementary measurements as a part of future work. For instance, when using SPAN-Ai measurements that are significantly outside the FOV, SPC measurements may be used to complement GDF reconstructions \citep{Kasper2016, Case_2020}. This is especially crucial to track changes, if any, to VDFs during switchbacks. Other avenues for future work involve exploring physics-based techniques to extrapolate the GDF outside the convex hull within which g-SBR performs the reconstructions. This is essential for being able to directly feed in \texttt{gdf} super-resolution outputs into plasma solvers such as ALPS \citep{Verscharen_ALPS_2018}. The effect of expanding the convex hull is another practical future investigation. Extending the convex hull would only directly affect the Cartesian approach. The broader implications of convex hull geometry on reconstruction accuracy will be addressed in future work.

Furthermore, the super-resolution results from g-SBR enable accurate measurements of higher order statistical moments, anisotropies, and measures of non-Maxwellianity, all essential quantities for quantifying the energy exchange in collisionless plasmas \citep{Richard_2025}. The g-SBR framework also provides a general methodology for characterizing the gyrotropic structure of any electrostatic analyzer measurement. This capability is particularly relevant to upcoming multi-spacecraft missions such as Helioswarm, where understanding the detailed phase-space structure of particle distributions will be critical for interpreting cross-scale energy transfer and kinetic coupling across the turbulent cascade \citep{klein2023helioswarm}.

\section*{acknowledgements}
    SBD and MT were supported by the Parker Solar Probe mission SWEAP investigation under NASA contract NNN06AA01C. SBD and MT are particularly grateful to Prof.~Frederik J.~Simons for numerous discussions and the MATLAB software used in the generation of the Slepian functions. These may be found in the publicly available packages released on Zenodo as \texttt{slepian\_alpha} \citep{slepian_alpha} and \texttt{slepian\_foxtrot} \citep{slepian_foxtrot}. Additionally, the authors thank Dr.~Michael L.~Stevens for his guidance and detailed feedback in preparation of this manuscript. We acknowledge Dr.~Kristoff Paulson, Prof.~Kristopher G.~Klein and Dr.~Davin Larson for critical input and detailed discussions throughout our study. 

    \software{
    Matlab \citep{MATLAB2025a}, 
    PySPEDAS \citep{pyspedas},
    emcee \citep{emcee}, 
    corner \citep{corner},
    NumPy \citep{numpy}, 
    SciPy \citep{scipy}, 
    Astropy \citep{astropy_v5},
    Matplotlib \citep{matplotlib},
    scikit-learn \citep{scikit-learn},
    Sphinx \citep{sphinx_doc},
    PlasmaPy \citep{PlasmaPy_Community_2024_PlasmaPy}, 
    Xarray \citep{xarray_v2025_4_0}
    }

\appendix

\section{B-spline regularization matrix} \label{sec:B-spline-Reg-matrix}

As mentioned in Section~\ref{sec:gyro-centroid-finder}, we use a default knot spacing which is approximately the log-spacing of grids in the energy dimension and angular degree resolution $L_{\mathrm{max}} = 12$ which is the Nyquist limit when the magnetic field is aligned with the radial direction of the instrument polar coordinates. However, the direction of $\boldsymbol{\hat{b}}$ causes the FAC grid distributions to change. In the polar-cap method, this can potentially result in large fluctuations due to under-constrained Slepian functions located on certain B-spline knots. Clearly such fluctuations, by virtue of a poorly conditioned inverse problem, is not physical. Therefore, we implement a smoothness regularization in the B-splines in $v$. In practice, this is done by penalizing large second derivatives as a function of $v$. Therefore, we introduce the following cost function
\begin{equation}
    \chi^2_{\mathrm{reg}} = \mu \iiint \left|\frac{\partial^2 f}{\partial v^2} \right|^2 \, v^2 \, \sin{\theta} \, \mathrm{d}\theta \, \mathrm{d}\phi \, \mathrm{d}v  = 2\pi \mu \iint \left|\frac{\partial^2 f}{\partial v^2} \right|^2 \, v^2 \, \sin{\theta} \, \mathrm{d}\theta \, \mathrm{d}v
\end{equation}
Now, for the polar-cap method, we parameterize the GDF as in Eqn.~(\ref{eqn:f_gyro_1DSlep}). Using this, the regularization cost function becomes
\begin{equation}
    \chi^2_{\mathrm{reg}} = 2\pi \mu c^{i_1 \alpha_1} \left(\int_{v_{\mathrm{min}}}^{v_{\mathrm{max}}} \beta''_{i_1}(v) \, \beta''_{i_2}(v) \, v^2 \mathrm{d}v \right) \, \left(\int_0^{\Theta} g_{\alpha_1}(\theta) \, g_{\alpha_2}(\theta) \, \sin{\theta} \, \mathrm{d}\theta \right) c^{i_2 \alpha_2} \, .
\end{equation}
By construction, the polar-cap Slepian functions are orthogonal. We may choose to write $\mathcal{N}_{\alpha_1} \, \delta_{\alpha_1, \alpha_2} = \int_0^{\Theta} g_{\alpha_1}(\theta) \, g_{\alpha_2}(\theta) \, \sin{\theta} \, \mathrm{d}\theta$ and $\mathcal{B}_{i_1 i_2} = \int_{v_{\mathrm{min}}}^{v_{\mathrm{max}}} \beta''_{i_1}(v) \, \beta''_{i_2}(v) \, v^2 \mathrm{d}v$. The above expression then simplifies to
\begin{equation}
       \chi^2_{\mathrm{reg}} = 2\pi \mu c^{i_1 \alpha_1} \mathcal{B}_{i_1, i_2} \, \mathcal{N}_{\alpha_1} \, \delta_{\alpha_1, \alpha_2} \, c^{i_2 \alpha_2} \, . 
\end{equation}
Finally, minimizing the cost function with respect to the coefficients $c^{i, \alpha}$, we get
\begin{equation}
    \frac{\partial \chi^2_{\mathrm{reg}}}{\partial c^{i \alpha}} = 4\pi \mu \,  c^{i_1\alpha} \mathcal{B}_{i_1, i} \, \mathcal{N}_{\alpha} \, .
\end{equation}
Performing the same index transformation as in Section~\ref{sec:polcap}, we go to the composite index $(i, \alpha) \rightarrow k$ and $(i_1, \alpha) \rightarrow k_1$, the final expression looks like
\begin{equation}
    \frac{\partial \chi^2_{\mathrm{reg}}}{\partial c^{i \alpha}} = 4\pi \mu \, c^{k_1} \, D_{k_1,k} \, \quad \mathrm{where,} \quad D_{k_1, k} \equiv \beta_{i_1,i} \, \mathcal{N}_{\alpha} \quad .
\end{equation}
We reference this as the regularization matrix $\mathbf{D}$ in Section~\ref{sec:hybrid}.

% \bibliography{references}{}

\begin{thebibliography}{}
\expandafter\ifx\csname natexlab\endcsname\relax\def\natexlab#1{#1}\fi
\providecommand{\url}[1]{\href{#1}{#1}}
\providecommand{\dodoi}[1]{doi:~\href{http://doi.org/#1}{\nolinkurl{#1}}}
\providecommand{\doeprint}[1]{\href{http://ascl.net/#1}{\nolinkurl{http://ascl.net/#1}}}
\providecommand{\doarXiv}[1]{\href{https://arxiv.org/abs/#1}{\nolinkurl{https://arxiv.org/abs/#1}}}

\bibitem[{{Astropy Collaboration} {et~al.}(2022){Astropy Collaboration},
  {Price-Whelan}, {Lim}, {Earl}, {Starkman}, {Bradley}, {Shupe}, {Patil},
  {Corrales}, {Brasseur}, {N{\"o}the}, {Donath}, {Tollerud}, {Morris},
  {Ginsburg}, {Vaher}, {Weaver}, {Tocknell}, {Jamieson}, {van Kerkwijk},
  {Robitaille}, {Merry}, {Bachetti}, {G{\"u}nther}, {Aldcroft},
  {Alvarado-Montes}, {Archibald}, {B{\'o}di}, {Bapat}, {Barentsen},
  {Baz{\'a}n}, {Biswas}, {Boquien}, {Burke}, {Cara}, {Cara}, {Conroy},
  {Conseil}, {Craig}, {Cross}, {Cruz}, {D'Eugenio}, {Dencheva}, {Devillepoix},
  {Dietrich}, {Eigenbrot}, {Erben}, {Ferreira}, {Foreman-Mackey}, {Fox},
  {Freij}, {Garg}, {Geda}, {Glattly}, {Gondhalekar}, {Gordon}, {Grant},
  {Greenfield}, {Groener}, {Guest}, {Gurovich}, {Handberg}, {Hart},
  {Hatfield-Dodds}, {Homeier}, {Hosseinzadeh}, {Jenness}, {Jones}, {Joseph},
  {Kalmbach}, {Karamehmetoglu}, {Ka{\l}uszy{\'n}ski}, {Kelley}, {Kern},
  {Kerzendorf}, {Koch}, {Kulumani}, {Lee}, {Ly}, {Ma}, {MacBride}, {Maljaars},
  {Muna}, {Murphy}, {Norman}, {O'Steen}, {Oman}, {Pacifici}, {Pascual},
  {Pascual-Granado}, {Patil}, {Perren}, {Pickering}, {Rastogi}, {Roulston},
  {Ryan}, {Rykoff}, {Sabater}, {Sakurikar}, {Salgado}, {Sanghi}, {Saunders},
  {Savchenko}, {Schwardt}, {Seifert-Eckert}, {Shih}, {Jain}, {Shukla}, {Sick},
  {Simpson}, {Singanamalla}, {Singer}, {Singhal}, {Sinha}, {Sip{\H{o}}cz},
  {Spitler}, {Stansby}, {Streicher}, {{\v{S}}umak}, {Swinbank}, {Taranu},
  {Tewary}, {Tremblay}, {de Val-Borro}, {Van Kooten}, {Vasovi{\'c}}, {Verma},
  {de Miranda Cardoso}, {Williams}, {Wilson}, {Winkel}, {Wood-Vasey}, {Xue},
  {Yoachim}, {Zhang}, {Zonca}, \& {Astropy Project Contributors}}]{astropy_v5}
{Astropy Collaboration}, {Price-Whelan}, A.~M., {Lim}, P.~L., {et~al.} 2022,
  \apj, 935, 167, \dodoi{10.3847/1538-4357/ac7c74}

\bibitem[{Bharati~Das \& Terres(2025)}]{Das_Terres_2025}
Bharati~Das, S., \& Terres, M. 2025, The Astrophysical Journal, 982, 96,
  \dodoi{10.3847/1538-4357/adb6a0}

\bibitem[{Bowen {et~al.}(2024)Bowen, Vasko, Bale, Chandran, Chasapis, Dudok~de
  Wit, Mallet, McManus, Meyrand, Pulupa, \& Squire}]{Bowen_2024}
Bowen, T.~A., Vasko, I.~Y., Bale, S.~D., {et~al.} 2024, The Astrophysical
  Journal Letters, 972, L8, \dodoi{10.3847/2041-8213/ad6b2e}

\bibitem[{Case {et~al.}(2020)Case, Kasper, Stevens, Korreck, Paulson, Daigneau,
  Caldwell, Freeman, Henry, Klingensmith, Bookbinder, Robinson, Berg, Tiu,
  Wright, Reinhart, Curtis, Ludlam, Larson, Whittlesey, Livi, Klein, \&
  Martinović}]{Case_2020}
Case, A.~W., Kasper, J.~C., Stevens, M.~L., {et~al.} 2020, The Astrophysical
  Journal Supplement Series, 246, 43, \dodoi{10.3847/1538-4365/ab5a7b}

\bibitem[{Chandran {et~al.}(2013)Chandran, Verscharen, Quataert, Kasper,
  Isenberg, \& Bourouaine}]{Chandran_2013}
Chandran, B. D.~G., Verscharen, D., Quataert, E., {et~al.} 2013, The
  Astrophysical Journal, 776, 45, \dodoi{10.1088/0004-637X/776/1/45}

\bibitem[{{Fazakerley} {et~al.}(1998){Fazakerley}, {Schwartz}, \&
  {Paschmann}}]{Fazakerley_Measurements}
{Fazakerley}, A.~N., {Schwartz}, S.~J., \& {Paschmann}, G. 1998, ISSI
  Scientific Reports Series, 1, 91

\bibitem[{{Foreman-Mackey}(2016)}]{corner}
{Foreman-Mackey}, D. 2016, The Journal of Open Source Software, 1, 24,
  \dodoi{10.21105/joss.00024}

\bibitem[{{Foreman-Mackey} {et~al.}(2013){Foreman-Mackey}, {Hogg}, {Lang}, \&
  {Goodman}}]{emcee}
{Foreman-Mackey}, D., {Hogg}, D.~W., {Lang}, D., \& {Goodman}, J. 2013, \pasp,
  125, 306, \dodoi{10.1086/670067}

\bibitem[{{Grimes} {et~al.}(2022){Grimes}, {Harter}, {Hatzigeorgiu}, {Drozdov},
  {Lewis}, {Angelopoulos}, {Cao}, {Chu}, {Hori}, {Matsuda}, {Jun}, {Nakamura},
  {Kitahara}, {Segawa}, {Miyoshi}, \& {Le Contel}}]{pyspedas}
{Grimes}, E.~W., {Harter}, B., {Hatzigeorgiu}, N., {et~al.} 2022, Frontiers in
  Astronomy and Space Sciences, 9, 1020815, \dodoi{10.3389/fspas.2022.1020815}

\bibitem[{Harris {et~al.}(2020)Harris, Millman, van~der Walt, Gommers,
  Virtanen, Cournapeau, Wieser, Taylor, Berg, Smith, Kern, Picus, Hoyer, van
  Kerkwijk, Brett, Haldane, del R{\'{i}}o, Wiebe, Peterson,
  G{\'{e}}rard-Marchant, Sheppard, Reddy, Weckesser, Abbasi, Gohlke, \&
  Oliphant}]{numpy}
Harris, C.~R., Millman, K.~J., van~der Walt, S.~J., {et~al.} 2020, Nature, 585,
  357, \dodoi{10.1038/s41586-020-2649-2}

\bibitem[{Howes(2024)}]{Howes_2024}
Howes, G.~G. 2024, Journal of Plasma Physics, 90, 905900504,
  \dodoi{10.1017/S0022377824001090}

\bibitem[{Howes {et~al.}(2006)Howes, Cowley, Dorland, Hammett, Quataert, \&
  Schekochihin}]{Howes_2006}
Howes, G.~G., Cowley, S.~C., Dorland, W., {et~al.} 2006, The Astrophysical
  Journal, 651, 590, \dodoi{10.1086/506172}

\bibitem[{Hoyer {et~al.}(2016)Hoyer, Fitzgerald, Hamman,
  {et~al.}}]{xarray_v2025_4_0}
Hoyer, S., Fitzgerald, C., Hamman, J., {et~al.} 2016, xarray: v2025.4.0,
  \dodoi{10.5281/zenodo.15306938}

\bibitem[{Hunter(2007)}]{matplotlib}
Hunter, J.~D. 2007, Computing in Science \& Engineering, 9, 90,
  \dodoi{10.1109/MCSE.2007.55}

\bibitem[{Kasper {et~al.}(2016)Kasper, Abiad, Austin, Balat-Pichelin, Bale,
  Belcher, Berg, Bergner, Berthomier, Bookbinder, Brodu, Caldwell, Case,
  Chandran, Cheimets, Cirtain, Cranmer, Curtis, Daigneau, Dalton, Dasgupta,
  DeTomaso, Diaz-Aguado, Djordjevic, Donaskowski, Effinger, Florinski, Fox,
  Freeman, Gallagher, Gary, Gauron, Gates, Goldstein, Golub, Gordon, Gurnee,
  Guth, Halekas, Hatch, Heerikuisen, Ho, Hu, Johnson, Jordan, Korreck, Larson,
  Lazarus, Li, Livi, Ludlam, Maksimovic, McFadden, Marchant, Maruca, McComas,
  Messina, Mercer, Park, Peddie, Pogorelov, Reinhart, Richardson, Robinson,
  Rosen, Skoug, Slagle, Steinberg, Stevens, Szabo, Taylor, Tiu, Turin, Velli,
  Webb, Whittlesey, Wright, Wu, \& Zank}]{Kasper2016}
Kasper, J.~C., Abiad, R., Austin, G., {et~al.} 2016, Space Science Reviews,
  204, 131, \dodoi{10.1007/s11214-015-0206-3}

\bibitem[{Klein {et~al.}(2021)Klein, Verniero, Alterman, Bale, Case, Kasper,
  Korreck, Larson, Lichko, Livi, McManus, Martinović, Rahmati, Stevens, \&
  Whittlesey}]{Klein_2021}
Klein, K.~G., Verniero, J.~L., Alterman, B., {et~al.} 2021, The Astrophysical
  Journal, 909, 7, \dodoi{10.3847/1538-4357/abd7a0}

\bibitem[{Klein {et~al.}(2023)Klein, Spence, Alexandrova, Argall, Arzamasskiy,
  Bookbinder, Broeren, Caprioli, Case, Chandran,
  {et~al.}}]{klein2023helioswarm}
Klein, K.~G., Spence, H., Alexandrova, O., {et~al.} 2023, Space Science
  Reviews, 219, 74

\bibitem[{{Laker} {et~al.}(2024){Laker}, {Horbury}, {Woodham}, {Bale}, \&
  {Matteini}}]{Laker2024}
{Laker}, R., {Horbury}, T.~S., {Woodham}, L.~D., {Bale}, S.~D., \& {Matteini},
  L. 2024, \mnras, 527, 10440, \dodoi{10.1093/mnras/stad3351}

\bibitem[{{Livi} {et~al.}(2022){Livi}, {Larson}, {Kasper}, {Abiad}, {Case},
  {Klein}, {Curtis}, {Dalton}, {Stevens}, {Korreck}, {Ho}, {Robinson}, {Tiu},
  {Whittlesey}, {Verniero}, {Halekas}, {McFadden}, {Marckwordt}, {Slagle},
  {Abatcha}, {Rahmati}, \& {McManus}}]{Livi_etal_2022}
{Livi}, R., {Larson}, D.~E., {Kasper}, J.~C., {et~al.} 2022, \apj, 938, 138,
  \dodoi{10.3847/1538-4357/ac93f5}

\bibitem[{Martinović {et~al.}(2020)Martinović, Klein, Kasper, Case, Korreck,
  Larson, Livi, Stevens, Whittlesey, Chandran, Alterman, Huang, Chen, Bale,
  Pulupa, Malaspina, Bonnell, Harvey, Goetz, Dudok~de Wit, \&
  MacDowall}]{Martinovic_2020}
Martinović, M.~M., Klein, K.~G., Kasper, J.~C., {et~al.} 2020, The
  Astrophysical Journal Supplement Series, 246, 30,
  \dodoi{10.3847/1538-4365/ab527f}

\bibitem[{{Paschmann} {et~al.}(1998){Paschmann}, {Fazakerley}, \&
  {Schwartz}}]{Paschmann_Moments}
{Paschmann}, G., {Fazakerley}, A.~N., \& {Schwartz}, S.~J. 1998, ISSI
  Scientific Reports Series, 1, 125

\bibitem[{Pedregosa {et~al.}(2011)Pedregosa, Varoquaux, Gramfort, Michel,
  Thirion, Grisel, Blondel, Prettenhofer, Weiss, Dubourg, Vanderplas, Passos,
  Cournapeau, Brucher, Perrot, \& Duchesnay}]{scikit-learn}
Pedregosa, F., Varoquaux, G., Gramfort, A., {et~al.} 2011, Journal of Machine
  Learning Research, 12, 2825

\bibitem[{{PlasmaPy Community et al.}(2024)}]{PlasmaPy_Community_2024_PlasmaPy}
{PlasmaPy Community et al.} 2024, PlasmaPy, version 2024.10.0, 2024.10.0,
  Zenodo, \dodoi{10.5281/zenodo.14010450}

\bibitem[{Richard {et~al.}(2025)Richard, Servidio, Svenningsson, Artemyev,
  Klein, Yordanova, Chasapis, Pezzi, \& Khotyaintsev}]{Richard_2025}
Richard, L., Servidio, S., Svenningsson, I., {et~al.} 2025, Phys. Rev. E, 112,
  L053201, \dodoi{10.1103/7p1x-84y9}

\bibitem[{Shankarappa {et~al.}(2023)Shankarappa, Klein, \&
  Martinović}]{Shankarappa_2023}
Shankarappa, N., Klein, K.~G., \& Martinović, M.~M. 2023, The Astrophysical
  Journal, 946, 85, \dodoi{10.3847/1538-4357/acb542}

\bibitem[{Simons(2020)}]{slepian_foxtrot}
Simons, F.~J. 2020, csdms-contrib/slepian\_foxtrot: 1.0.1, 1.0.1,  Zenodo,
  \dodoi{10.5281/zenodo.4085250}

\bibitem[{{Simons} \& {Dahlen}(2006)}]{Slepian_polar_caps}
{Simons}, F.~J., \& {Dahlen}, F.~A. 2006, Geophysical Journal International,
  166, 1039, \dodoi{10.1111/j.1365-246X.2006.03065.x}

\bibitem[{Simons {et~al.}(2020)Simons, Harig, Plattner, von Hippel, \&
  Albert}]{slepian_alpha}
Simons, F.~J., Harig, C., Plattner, A., von Hippel, M., \& Albert. 2020,
  csdms-contrib/slepian\_alpha: Release 1.0.5, 1.0.5,  Zenodo,
  \dodoi{10.5281/zenodo.4085210}

\bibitem[{{Simons} \& {Wang}(2011)}]{Simons&Wang2011}
{Simons}, F.~J., \& {Wang}, D.~V. 2011, GEM - International Journal on
  Geomathematics, 2, 1, \dodoi{10.1007/s13137-011-0016-z}

\bibitem[{{Sphinx Developers}(2025)}]{sphinx_doc}
{Sphinx Developers}. 2025, Sphinx Documentation, Python Software Foundation.
\newblock \url{https://www.sphinx-doc.org/en/master/}

\bibitem[{Terres \& Li(2022)}]{Terres_2022}
Terres, M., \& Li, G. 2022, The Astrophysical Journal, 924, 53,
  \dodoi{10.3847/1538-4357/ac400c}

\bibitem[{{The MathWorks Inc.}(2025)}]{MATLAB2025a}
{The MathWorks Inc.} 2025, {MATLAB} version: {R2025a},  Natick, Massachusetts,
  United States: {The MathWorks Inc.}
\newblock \url{https://www.mathworks.com}

\bibitem[{Vech {et~al.}(2018)Vech, Mallet, Klein, \& Kasper}]{Vech_2018}
Vech, D., Mallet, A., Klein, K.~G., \& Kasper, J.~C. 2018, The Astrophysical
  Journal Letters, 855, L27, \dodoi{10.3847/2041-8213/aab351}

\bibitem[{Verniero {et~al.}(2020)Verniero, Larson, Livi, Rahmati, McManus,
  Pyakurel, Klein, Bowen, Bonnell, Alterman, Whittlesey, Malaspina, Bale,
  Kasper, Case, Goetz, Harvey, Korreck, MacDowall, Pulupa, Stevens, \&
  de~Wit}]{Verniero_2020}
Verniero, J.~L., Larson, D.~E., Livi, R., {et~al.} 2020, The Astrophysical
  Journal Supplement Series, 248, 5, \dodoi{10.3847/1538-4365/ab86af}

\bibitem[{Verscharen {et~al.}(2018)Verscharen, Klein, Chandran, Stevens, Salem,
  \& Bale}]{Verscharen_ALPS_2018}
Verscharen, D., Klein, K., Chandran, B., {et~al.} 2018, Journal of Plasma
  Physics, 84, 905840403, \dodoi{10.1017/S0022377818000739}

\bibitem[{Verscharen {et~al.}(2019)Verscharen, Klein, \&
  Maruca}]{verscharen2019multi}
Verscharen, D., Klein, K.~G., \& Maruca, B.~A. 2019, Living Reviews in Solar
  Physics, 16, 5, \dodoi{10.1007/s41116-019-0021-0}

\bibitem[{Virtanen {et~al.}(2020)Virtanen, Gommers, Oliphant, Haberland, Reddy,
  Cournapeau, Burovski, Peterson, Weckesser, Bright, {van der Walt}, Brett,
  Wilson, Millman, Mayorov, Nelson, Jones, Kern, Larson, Carey, Polat, Feng,
  Moore, {VanderPlas}, Laxalde, Perktold, Cimrman, Henriksen, Quintero, Harris,
  Archibald, Ribeiro, Pedregosa, {van Mulbregt}, \& {SciPy 1.0
  Contributors}}]{scipy}
Virtanen, P., Gommers, R., Oliphant, T.~E., {et~al.} 2020, Nature Methods, 17,
  261, \dodoi{10.1038/s41592-019-0686-2}

\bibitem[{Woodham {et~al.}(2021)Woodham, Horbury, Matteini, Woolley, Laker,
  Bale, Nicolaou, Stawarz, Stansby, Hietala, {et~al.}}]{Woodham_2021}
Woodham, L., Horbury, T., Matteini, L., {et~al.} 2021, Astronomy \&
  Astrophysics, 650, L1, \dodoi{10.1051/0004-6361/202039415}

\end{thebibliography}
% \bibliographystyle{aasjournal}

\end{document}